\documentclass[twocolumn]{aastex631}

\hypersetup{
   colorlinks,
   linkcolor={blue!88!black!80},
   citecolor={blue!88!black!80},
   urlcolor={blue!88!black!80}}

\usepackage{float}
\usepackage{mathrsfs}
\usepackage{natbib}
\usepackage{xspace}
\usepackage{hyperref}
\usepackage{multirow}
\usepackage{booktabs}
\usepackage{float}

\newcommand{\RNum}[1]{\uppercase\expandafter{\romannumeral #1\relax}}

\newcommand{\hbeta}{H{$\beta$}}
\newcommand{\halpha}{H{$\alpha$}}

\def\FeII{Fe\,{\sc ii}}

\def\HeIIopt{He\,{\sc ii}\,$\lambda$4687}

\def \OIII {[O\,{\sc iii}]}
\newcommand{\OIIIa}{[O{\sevenrm\,III}]\,$\lambda$4959}
\newcommand{\OIIIb}{[O{\sevenrm\,III}]\,$\lambda$5007}

\newcommand{\NIIab}{[N\,{\sevenrm\,II}]\,$\lambda\lambda$6548,6584}

\newcommand{\SII}{[S{\sevenrm\,II}]}

\newcommand{\SIIab}{[S\,{\sevenrm\,II}]\,$\lambda\lambda$6717,6731}

   \font\sevenrm=cmr7 scaled 1000
\newcommand{\comments}[1]{}

\begin{document}

\title{Identifying changing-look AGNs using variability characteristics}

\author[0000-0002-2052-6400]{Shu Wang}
\affiliation{Department of Physics \& Astronomy, Seoul National University, Seoul 08826, Republic of Korea; wangshu100002@gmail.com; woo@astro.snu.ac.kr}

\author[0000-0002-8055-5465]{Jong-Hak Woo}
\affiliation{Department of Physics \& Astronomy, Seoul National University, Seoul 08826, Republic of Korea; wangshu100002@gmail.com; woo@astro.snu.ac.kr}

\author[0000-0002-8055-5465]{Elena Gallo}
\affiliation{Department of Astronomy, University of Michigan, Ann Arbor, MI 48109, USA}

\author[0000-0001-8416-7059]{Hengxiao Guo}
\affiliation{Key Laboratory for Research in Galaxies and Cosmology, Shanghai Astronomical Observatory, Chinese Academy of Sciences, 80 Nandan Road, Shanghai 200030, People's Republic of China; hengxiaoguo@gmail.com}

\author[0000-0002-8055-5465]{Donghoon Son}
\affiliation{Department of Physics \& Astronomy, Seoul National University, Seoul 08826, Republic of Korea; wangshu100002@gmail.com; woo@astro.snu.ac.kr}


\author{Minzhi Kong}
\affiliation{Department of Physics, Hebei Normal University, No. 20 East of South 2nd Ring Road, Shijiazhuang 050024, People’s Republic of China}

\author[0000-0002-8055-5465]{Amit Kumar Mandal}
\affiliation{Department of Physics \& Astronomy, Seoul National University, Seoul 08826, Republic of Korea; wangshu100002@gmail.com; woo@astro.snu.ac.kr}

\author[0000-0002-8055-5465]{Hojin Cho}
\affiliation{Department of Physics \& Astronomy, Seoul National University, Seoul 08826, Republic of Korea; wangshu100002@gmail.com; woo@astro.snu.ac.kr}

\author[0000-0002-8055-5465]{Changseok Kim}
\affiliation{Department of Physics \& Astronomy, Seoul National University, Seoul 08826, Republic of Korea; wangshu100002@gmail.com; woo@astro.snu.ac.kr}

\author[0000-0002-8055-5465]{Jaejin Shin}
\affiliation{Korea Astronomy and Space Science Institute, Daejeon 34055, Republic of Korea}
\affiliation{Major in Astronomy and Atmospheric Sciences, Kyungpook National University, Daegu 41566, Republic of Korea}

\begin{abstract}
Changing-look (CL) Active Galactic Nuclei (AGNs), characterized by appearance/disappearance of broad emission lines in the span of a few years, present a challenge for the AGN unified model, whereby the Type 1 vs. Type 2 dichotomy results from orientation effects alone. We present a systematic study of a large sample of spectroscopically classified AGNs, using optical variability data from the Zwicky Transient Facility (ZTF) as well as follow-up spectroscopy data. We demonstrate that Type 1 vs. 2 AGN can be neatly separated on the basis of the variability metric $\sigma_{\rm QSO}$, which quantifies the resemblance of a light curve to a damp random walk model. For a small sub-sample, however, the ZTF light curves are inconsistent with their previous classification, suggesting the occurrence of a CL event. Specifically, we identify 35 (12) turn-on (turn-off) CL AGN candidates at $z<0.35$. Based on follow-up spectroscopy, we confirm 17 (4) turn-on (turn-off) CL~AGNs out of 21 (5) candidates, presenting a high success rate of our method. Our results suggest that the occurrence rate of CL~AGNs is $\sim$0.3\% over timescales of 5 to 20 years, and confirm that the CL transition typically occurs at the Eddington ratio of $\lesssim 0.01$.  

\end{abstract}


\section{Introduction} \label{sec:intro}

Active Galactic Nuclei (AGNs) are classified as either Type 1 or Type 2, respectively, based on the presence or absence of broad emission lines, which are typically defined with a full-width-at-half-maxima larger than $\sim$$1000$ km s$^{-1}$ in their ultra-violet (UV), optical, and near-IR spectra. Intermediate classes. i.e., Type 1.8 or 1.9 AGNs are characterized by broad \halpha\ and weak or absent broad \hbeta\ lines \citep{Osterbrock81}. In the AGN unified model, this dichotomy is purely driven by orientation effects \citep{Antonucci93,Urry95}, depending on whether or not the observer's line of sight is intercepted by a dusty torus.
This scenario is supported by multiple lines of evidence, including the discovery of broad lines in the polarized spectra of some Type 2 AGNs \citep{Antonucci85,Zakamska05} and the higher fraction of Compton-thick AGNs amongst Type 2 AGNs \citep{Mulchaey92}.

At the same time, the unified model faces significant challenges, particularly with the discovery of a small population of AGNs with  emerging/disappearing broad emission lines over timescales of several years. Approximately 150 such changing-look (CL) AGNs have been reported so far,  based on  
\hbeta\ and/or \halpha\ \citep[e.g.,][]{LaMassa15,Macleod16,Runnoe16,Ruan16,Runco16,Yang18,Macleod19,Wang-J19,Green22,Zeltyn22,Hon22, Navas22, Navas23b, Guo23Arxiv, Neustadt23}. These CL AGNs demonstrate that the orientation-based unified model might be too simplistic. 

The growing sample of CL AGNs provides insights to the physics underlying the phenomenon. CL AGNs are rare, with an occurrence rate of $\sim1$ per cent or lower, estimated using multi-epoch spectra from large-area spectroscopic surveys \citep[e.g.,][]{Macleod16,Macleod19,Yang18}. 
The timescale of CL events is approximately 1 to 10 years \citep[][and references therein]{Ricci22}, while in the most extreme cases, the broad emission lines can appear/disappear within $\sim$months \citep{Trakhtenbrot19a,Zeltyn22}.  Multi-wavelength observations reveal that the appearance of broad emission lines occurs nearly synchronously with mid-infrared, optical and X-ray continuum variations \citep{Sheng17, Temple23,Yang23}. Since these behaviors (in addition to the low polarization level observed in turned-off AGNs,  \citealt{Hutsemekers19}), can not be accounted for by changing obscuration, the driver(s) of the CL phenomenon is generally ascribed to global changes in accretion power \citep[e.g.,][]{Elitzur09,Elitzur14,Ricci17,Guo19,Guo20a}. However, the global change of accretion power ought to occur on viscous timescales, which are too long compared to the observed transitions. This has led to alternative suggestions such that CL events might be triggered by thermal instabilities, or that other processes (e.g., magnetically elevated accretion) ought to be involved in order to reduce the characteristic inflow timescale \citep{Dexter19,Sniegowska20,Pan21}. Lastly, tidal disruption events (TDEs), could be responsible for a small fraction of the observed CL AGNs \citep{Merloni15,Ricci20,Li22}. %

Many previous searches for CL~AGNs focus on extreme variable AGNs (EVAs), i.e., events showing large amplitude variability in $g$ band (with variations in excess of 1 mag; \citealt{Macleod19,Yang18}).  For those, the success rate of follow-up spectroscopic observations, i.e., the fraction of EVAs that are indeed associated with CL AGNs, is rather moderate, between 15$\sim$50\% (albeit the exact fraction is highly dependent on the chosen threshold and the signal-to-noise level;  \citealt{Macleod19,Guo20}).  

\begin{figure}[tbp]
    \centering
    \includegraphics[width=0.49\textwidth]{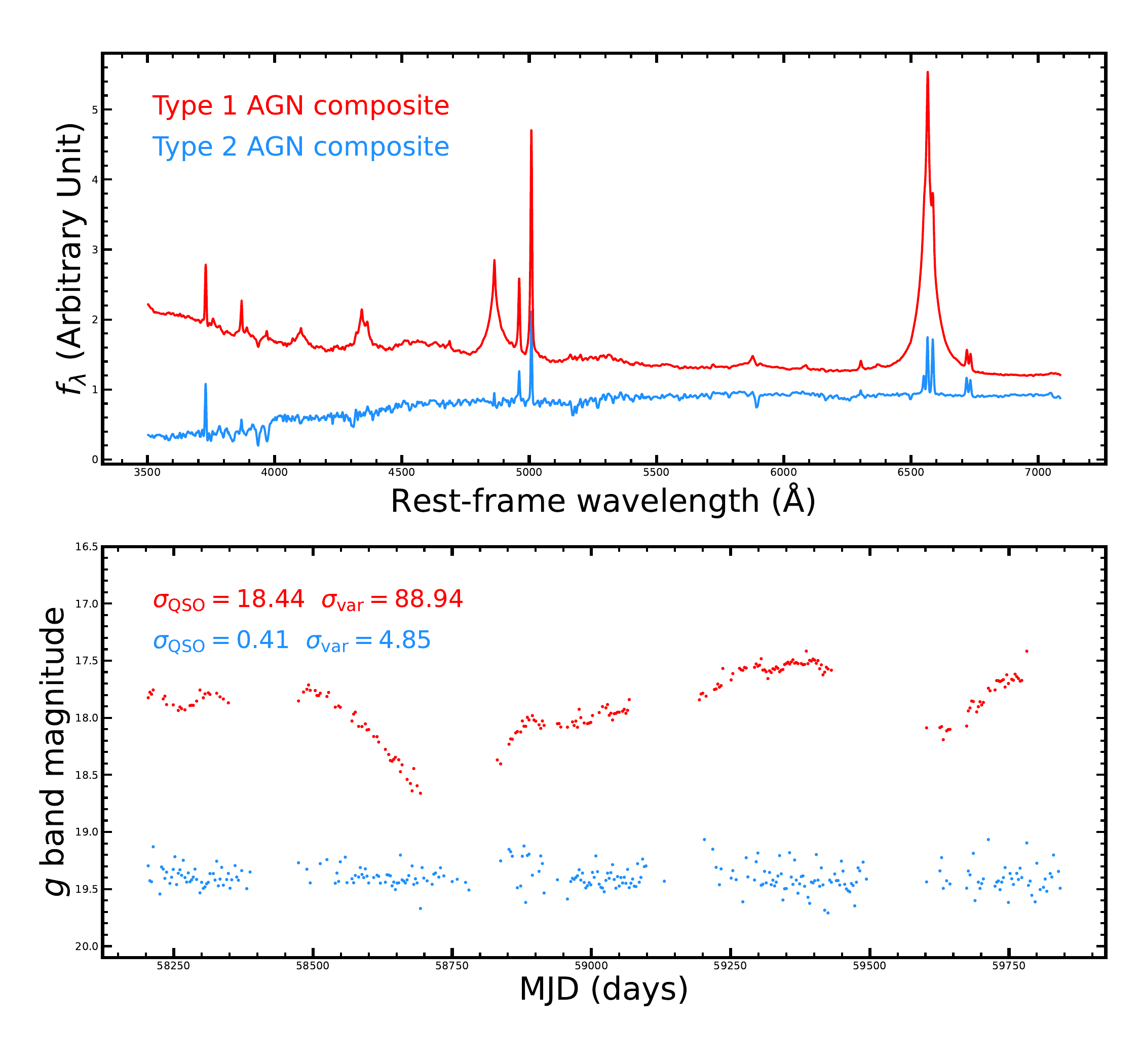}
    \caption{Composite spectrum of Type 1 (red) and Type 2 (blue) AGN, respectively (upper panel), generated using a parent sample with SDSS spectra. The Type 2 composite spectrum is shifted for demonstration purpose. The lower panel shows two examples of 3-day binned cleaned ZTF light curves for Type 1 (red) and 2  (blue) AGNs. $\sigma_{\rm QSO}$ and $\sigma_{\rm var}$ of these two light curves are labeled in upper left, which indicate the significance of the light curves to be described by the damped random walk model and the level of the variability, respectively (see \S \ref{sec:method} for details). }
    \label{fig:LC}
\end{figure}

A novel approach to distinguish between Type 1 and 2 AGNs is based on their variability pattern. As shown in Figure \ref{fig:LC}, Type 1 AGNs show stochastically variable light curves that can be statistically described by a damped random walk (DRW) model \citep[e.g.,][]{Kelly09,Macleod10}. In contrast, due to their highly obscured nature, the light curves of Type 2 AGNs are usually flat, and dominated by white-noise-like scatter \citep[e.g.,][]{Yip09,Barth14, Sanchez17, Navas23a}. 

Nevertheless, some spectroscopically classified Type 2 AGNs are known to exhibit some non-negligible levels of variability \citep{Barth14, Navas23a,Zhang23}, with an occurrence rate of $\sim10$\%. Some of these are likely mis-classified, weak, Type 1 AGNs, whereas others have been identified as CL~AGNs based on follow-up spectroscopic observations \citep[][hereafter LN22 and LN23b, respectively]{Navas22,Navas23b}.   

Large time-domain surveys offer an excellent opportunity to investigate the variability patterns of different AGN classes, and to identify bona fide CL~AGNs  \citep[e.g.,][]{Gezari17,Frederick19,Graham20,Sanchez21,Sanchez23,Navas22, Navas23a, Navas23b}. 
In this work, we perform a systematic variability study of a large sample of spectroscopically classified AGNs, using light curves from the Zwicky Transient Facility \citep[ZTF;][]{Bellm19}, and demonstrate the viability of this method for the purpose of identifying candidate CL AGNs with a high success rate.

The manuscript is organized as follow.   \S\ref{sec:method} presents the method for selecting CL AGN candidates based on variability properties. \S\ref{sec:observation} describes the follow-up spectroscopic observations;   \S\ref{sec:results} presents the results of the spectroscopic campaign.  In \S\ref{sec:discussion}, we compare our method with traditional selection approaches, and assess the typical timescales and Eddington ratios that are associated with confirmed CL AGNs. We also estimate the CL AGN occurrence rate based on our sample. \S\ref{sec:summary} summarizes our work and discusses prospects of upcoming time domain surveys for fully understanding the nature of CL AGNs. Throughout this paper, we use a flat cosmology with $h_0=70.0$ and $\Omega_0=0.3$.

\section{Candidate CL AGNs: Sample Selection} \label{sec:method}

Type 1 and 2 AGNs are known to exhibit different variability patterns. Our objective is to search for AGNs whose ZTF-measured light curves are inconsistent with their historical spectral types. 
ZTF is an on-going time domain survey which starts in 2018. It covers the entire northern sky and its public data release (DR) provides relatively deep \citep[$g$ band limiting magnitude 20.8 mag;][]{Masci19}, high-cadence ($\sim$3 days)  photometric light curves for large sample of AGNs. 

Specifically, we aim to isolate Type 1 [/Type 2] AGNs by selecting for high amplitude [/flat] variability light curves. These are commonly referred to as turn-on and turn-off CL AGN candidates, respectively.  

To identify candidate turn-on CL~AGNs, we start from Type 2 quasars/AGNs from the Million Quasar Catalog (MQC) v7.2 \citep{Flesch21}, and down-select for: (i) type/class 'K' (narrow-line Seyferts) and/or 'N' (narrow-line Quasi Stellar Objects; QSOs), and (ii) redshifts $z<0.35$, so as to ensure spectroscopic coverage in the \hbeta\ and \halpha\ (observed) wavelength windows. This yields a parent sample of $20\,980$ Type 2 AGNs.  Note that this Type 2 AGN parent sample may contain a small fraction of mis-classified weak Type 1 AGNs \citep[e.g.,][]{Navas23a}. Given the large sample size, we plan to check this potential contamination after the variability selection.

For candidate turn-off CL~AGNs, we start from the SDSS DR 14 Quasar Catalog \citep{Paris18}, which contains $526\,356$ visually confirmed quasars from SDSS-I, II, III, as well as pipeline-identified quasars from SDSS-IV through July 2016. 
We then refine our sample to quasars at $z<0.35$, aligning with the criteria of turn-on selection.
Based on \citet{Rakshit20}, we exclude AGNs with insufficiently broad H$\beta$ profiles, by requiring full-width-at-half-maxima (FWHM) larger than 1$\,$500 km s$^{-1}$.  We also exclude AGNs with red colors, by requiring $u-r<1.0$, as these objects can be intrinsically obscured, host-dominated (i.e., Type 1.9), or mis-classified.  The resulting Type 1 parent sample consists of $5\,067$ AGNs.  

The upper panel of Figure \ref{fig:LC} displays the Type 1 and 2 AGN composite spectra derived from the two parent samples described above. 

\begin{figure}[htbp]
    \centering
  \includegraphics[width=0.5\textwidth]{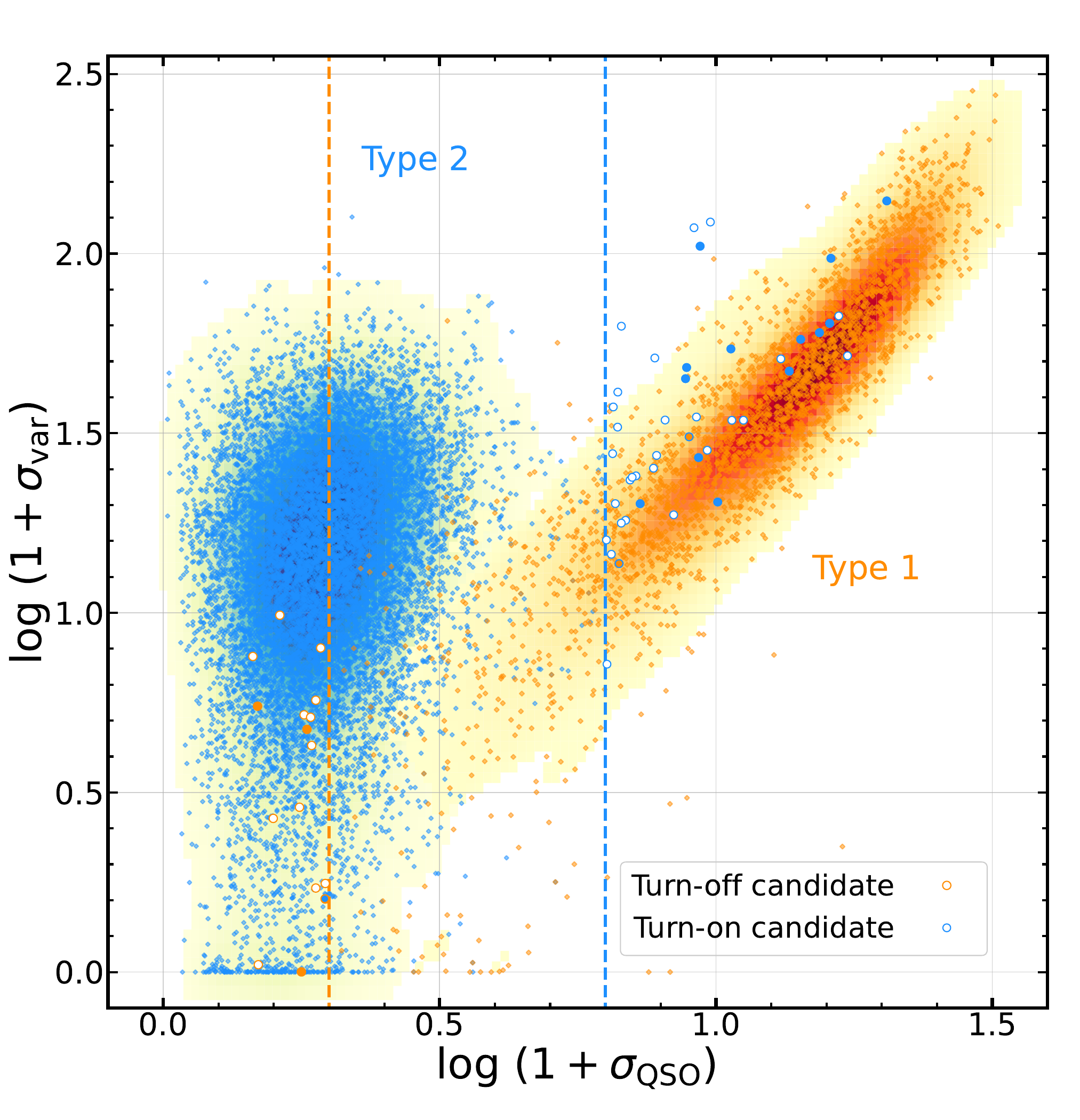} 
  \caption{Demonstration of the distinct distribution of Type 1 (orange) and Type 2 (blue) AGNs on the $\sigma_{\rm QSO}$--$\sigma_{\rm var}$ plane. The majority of Type 1 and 2 AGNs are well separated while a subset of both Type 1 and 2 AGNs are overlapped with locus of the opposite type, which we selected as our CL~AGN candidates. The blue (orange) dashed line refers to our criteria on $\sigma_{\rm QSO}$ for selecting turn-on(off) CL~AGNs (see text for details). }
    \label{fig:illustration}
\end{figure}

ZTF light curves (from 2018 to 2022) are acquired by cross-matching our Type 1 and 2 AGN parent samples with the ZTF public DR 15, via IRSA catalog query service\footnote{\url{https://irsa.ipac.caltech.edu/Missions/ztf.html}}. We adopt $g$ band light curves for our analysis because shorter wavelengths tend to exhibit larger variations. Further, we require at least 200 good epochs as defined by ZTF pipeline, so as to ensure high-cadence sampling over the four-year baseline. 

To classify a system as a candidate turn-off CL~AGNs, we require a median $g$ band magnitude brighter than 20.0 \citep[i.e., 0.8 magnitude brighter than the $g$ band detection limit 20.8;][]{Masci19} and an average magnitude uncertainty smaller than 0.2 mag. These requirements are meant to avoid contamination from objects with noisy light curves close to the detection limit.

Following LN22, we carry out several additional steps to clean the light curves; we only keep epochs with {\tt catflags} equal 0, which indicates no bad photometry. Second, we remove epochs with magnitude uncertainties more than twice the average uncertainties. Third, we perform a 3-$\sigma$ clipping procedure (twice) to remove outliers. Fourth, we perform a 3-day binning to reduce scatter.  These processes only slightly modify the number of epochs.  Figure \ref{fig:LC} shows two examples of cleaned, binned light curves, for a Type 1 and Type 2 AGN, respectively. Whereas the latter is flat and dominated by white-noise like scatter, the former is characterized by large, smoothly variable structures.

Next, we quantify the variability statistics using the {\tt qso\_fit}\footnote{\url{http://butler.lab.asu.edu/qso\_selection/index.html}} software \citep{Butler11}, which was developed to separate (Type 1) quasars from other types of point-like sources (e.g., stars) in SDSS Stripe 82 based on their $g$ band light curves. For each light curve, the software provides three distinct variability metrics; (i) $\sigma_{\rm var}$ assesses whether the source is variable, and the significance of said variability; (ii) $\sigma_{\rm QSO}$ quantifies to which extent the source variability is better described by a DRW model rather than a time-independent variable Gaussian signals; (iii) $\sigma_{\rm NotQSO}$ quantifies to which extent the source variability is well described by a time-independent Gaussian variable signals. Combined, these metrics have been proven to be very efficient at separating QSOs from stars with high completeness ($>99$\%) and purity ($<3$\% contamination) in SDSS Stripe 82.

Figure \ref{fig:illustration} confirms that the vast majority of Type 1 vs. Type 2 AGNs separate very well in the $\sigma_{\rm var}$--$\sigma_{\rm QSO}$ plane, with most Type 1s exhibiting high values of $\sigma_{\rm QSO}$ and $\sigma_{\rm var}$, and most Type 2s having low $\sigma_{\rm QSO}$ and moderate $\sigma_{\rm var}$. 

Whereas the measured difference in $\sigma_{\rm QSO}$ (between Type 1 and 2) is consistent with the indication from previous studies \citep{Barth14, Sanchez17}, the typical $\sigma_{\rm var}$ values for Type 2 AGNs are not consistent with zero; this likely due to the fact that the PSF-based photometry that we are using for extended sources is affected by variable seeing conditions \citep{Sanchez21}.  As a result of the cleaner separation, we thus opt to make use of $\sigma_{\rm QSO}$ to select candidate CL AGNs (we also tested $\sigma_{\rm NotQSO}$ but found that it did not provide any further insight).

We also discover a small subset of AGNs whose location in the $\sigma_{\rm QSO}$--$\sigma_{\rm var}$ plane does not match the original spectroscopic classification, indicating a possible CL event. 

\begin{table}[htbp]
\caption{Sample selection for turn-on CL~AGNs}\label{tab:Selection-on}
\begin{center}
\begin{tabular}{c l c}
\hline \hline
& Criterion & N$_{\rm object}$  \\ \hline 
  1 & Type 2 AGNs at $z<0.35$ from MQC & $20\,980$ \\
  2 & ZTF $g$ band $N_{\rm good\, epoch}>200$  & $9\,904$ \\
  3 & log (1+$\sigma_{\rm QSO}) > 0.8 $  & 42 \\
  4 & Visual Inspection to remove transients & 41 \\
  5 & Historical SNR$_{\rm BH\beta} < 2$ & 35 \\  
  \hline
\end{tabular}
\end{center}
\end{table}%

\begin{table}[htbp]
\caption{Sample selection for turn-off CL~AGNs}\label{tab:Selection-off}
\begin{center}
\begin{tabular}{c l c}
\hline \hline
  &  Criterion & N$_{\rm object}$  \\ \hline 
  1 &  Type 1 AGNs at $z<0.35$ from SDSS DR14  & $5\,067$ \\
                               & \& $u-r<1.0$ \& ${\rm FWHM}_{\rm H\beta}>1500$ km s$^{-1}$ & \\
  2 & ZTF $g$ band $N_{\rm good\, epoch}>200$  \& $g_{\rm MED} <$ 20.0 & $2\,988$ \\
                   &  \& $\overline{g_{\rm err}}<0.2$ & \\
  3 & log $(1+\sigma_{\rm QSO})< 0.3$ & 16  \\   
  4 & Visual inspection   & 12 \\
  \hline
\end{tabular}
\end{center}
\end{table}%

Specifically, we isolate 42 Type 2 AGNs (out of 9$\,$904 Type 2 AGNs with well-sampled ZTF light curves) with log (1+$\sigma_{\rm QSO})>0.8$  ($\sigma_{\rm QSO} \gtrsim 5$) as candidate turn-on CL~AGN. 
Visual inspection confirms that all but one system (which is likely a supernova or tidal disruption events) indeed exhibit textbook Type-1-like variability. 
Further details on the candidate turn-on CL AGN sample are provided in Appendix \ref{sec:appendixA}, where we use BPT diagnostics \citep{Baldwin81} to confirm that the SDSS spectra of all 41 systems are consistent with Type 2 AGNs rather than star-forming galaxies or low-ionization nuclear emission-line regions (LINERS). Last, we cull those AGNs whose historical spectra show detectable broad \hbeta\ profiles (see Figure \ref{fig:SNR_hbha}). The final sample includes 35 candidate turn-on CL AGNs.

Candidate turn-off CL AGNs are initially isolated as Type 1 AGNs with log (1+$\sigma_{\rm QSO} )< 0.3$, which results in 16 systems out of $2\,988$ objects in the parent sample that have well-sampled ZTF light curves. This threshold corresponds to $\sigma_{\rm QSO}<1$, which indicates that the observed light curves are more consistent with non-quasars rather than quasars. We visually remove 2 objects from our sample due to the presence of variability features in individual season. In addition, we exclude another 2 objects that have no prominent broad lines in their SDSS spectra. The final sample contains 12 candidate turn-off CL AGNs.  We note that the majority of Type 2 AGNs in Figure \ref{fig:illustration} have log $(1+\sigma_{\rm QSO})<0.5$ (i.e. $\sigma_{\rm QSO}\lesssim2$).
Based on a visual inspection, we find that this threshold includes light curves with small amplitude variability features, thus we elect to adopt this stricter threshold for selecting candidate turn-off CL~AGNs. 

The criteria for selecting (candidate) turn-on/off CL~AGN are summarized in Table \ref{tab:Selection-on} and \ref{tab:Selection-off}.  The final list of candidates turn-on and turn-off is summarized in Table \ref{tab:turn-on-candidates} and \ref{tab:turn-off-candidates}, respectively.

\section{Follow-up Spectroscopic Observations}\label{sec:observation}

\begin{table*}[htbp]
\caption{Summary of spectroscopic observation of 15 turn-on and 3 turn-off candidates}\label{tab:observation}
\begin{center}
\footnotesize
\begin{tabular}{c c c c c  c  c  c  c  c  c c c c}
\hline \hline
&   &    & \multicolumn{4}{c}{Historical observation}  & \multicolumn{6}{c}{New observation}  & \\
   \cmidrule(r){4-7} 
   \cmidrule(r){8-13}
& Object name  &  z  & MJD & Facility & SNR  & Type   & MJD & Facility & Instrument  &  $t_{\rm exp}$ (s) & SNR  & Type \\  \hline
   
 \multicolumn{13}{c}{Turn-on candidates} \\ \hline
1 & J081917.50+301935.6 & 0.0975 & 52619 & SDSS &  23.4  & Type 2   & 59685  & MDM       & OSMOS      & 3600     & 37.7  & Type 1 \\ 
2 & J091803.34+285815.9 & 0.2401 & 53388 & SDSS &  9.4    & Type 1.9   & 57724  & LAMOST & LRS      & \nodata & 8.3    & Type 1 \\ 
3 & J095137.27+341612.2 & 0.1322 & 53388 & SDSS &  10.2  & Type 2   & 59986  & MDM       & OSMOS     & 3600     & 13.3   & Type 1  \\ 
4 & J100906.06+353932.6 & 0.1101 & 53389 & SDSS &  25.3  & Type 1.9   & 59686  & MDM       & OSMOS      & 3600     & 36.8   & Type 1.9 \\ 
5 & J102038.50+243708.3 & 0.1894 & 53734 & SDSS &  13.1  & Type 1.9   & 60059  & MDM       & OSMOS      & 2700     & 27.8   & Type 1 \\ 
6 & J115000.56+350356.6 & 0.0611 & 53491 & SDSS &  24.9  & Type 1.9   & 59665  & MDM       & OSMOS      & 3600     & 32.2   & Type 1 \\ 
7 & J124617.33+282033.8 & 0.0995 & 54205 & SDSS &  26.2  & Type 1.9   & 60059  & MDM       & OSMOS      & 2700     & 22.3   & Type 1 \\ 
8 & J142352.08+245417.1 & 0.0744 & 53493 & SDSS &  19.1  & Type 1.9   & 59715  & MDM        & OSMOS      & 3600    & 22.3    & Type 1 \\ 
9 & J152406.53+432757.1 & 0.1984 & 52468 & SDSS &  8.9    & Type 2   & 58988  & LAMOST  & LRS     & \nodata & 9.2    & Type 1 \\ 
10 & J153832.66+460734.9 & 0.2026 & 52781 & SDSS &  9.5  & Type 2 & 59740  & Gemini-N  & GMOS  & 240     & 42.6    & Type 1 \\ 
11 & J154755.38+030350.8 & 0.0947 & 52045 & SDSS &  15.8  & Type 1.9  & 60151  & Gemini-N & GMOS  & 600     & 74.2  & Type 1 \\ 
12 & J155259.94+210246.8 & 0.1717 & 53557 & SDSS &  7.1    & Type 2  & 59736 & Gemini-N  & GMOS  & 540     & 77.5   & Type 1 \\ 
13 & J161219.56+462942.5 & 0.1254 & 52443 & SDSS &  14.0  & Type 2  & 59740 & Gemini-N  & GMOS  & 180     & 34.5   & Type 1  \\ 
14 & J163639.57+194201.7 & 0.1496 & 53224 & SDSS &  12.4  & Type 2  & 60151 & Gemini-N  & GMOS  & 600     & 39.8   & Type 2 \\ 
15 & J222559.66+201944.7 & 0.1554 & 56189 & BOSS &  17.0  & Type 1.9  & 59808 & Gemini-N  & GMOS  & 180     & 34.2   & Type 1 \\ \hline

\multicolumn{13}{c}{Turn-off candidates} \\ \hline
1 & J123020.67+462745.7  & 0.1803 & 53084 & SDSS &  16.4  & Type 1  & 56310 & LAMOST  & LRS     & \nodata & 7.7   & Type 1 \\ 
   &                                       &             &            &            &           &            & 58546 & LAMOST  & LRS      & \nodata & 11.3  & Type 1.9 \\ 
2 & J133241.17+300106.5  & 0.2204 & 53467 & SDSS &  14.8  & Type 1 & 57778 & LAMOST  & LRS     & \nodata & 7.6   & Type 1.9 \\ 
   &                                       &             &            &            &           &            & 60154 & Gemini-N  & GMOS  & 600      & 38.6   & Type 1.9\\
3 &J140007.94+275133.3   & 0.3338 & 53852& SDSS  &  14.0  & Type 1  & 60161 & Gemini-N  & GMOS  & 360      & 30.2   & Type 1 \\  \hline
\end{tabular}

\end{center}
\end{table*}

\begin{table*}[htbp]
\caption{Summary of literature observation of 6 turn-on and 2 turn-off  CL AGNs candidates}\label{tab:additional_observation}
\begin{center}
\footnotesize
\begin{tabular}{c c c c c  c   c  c  c}
\hline \hline
& &    & \multicolumn{2}{c}{Historical observation}  & \multicolumn{2}{c}{New observation}  & \\
   \cmidrule(r){4-5} 
   \cmidrule(r){6-7}
& Object name  &  z  & MJD & Facility  & MJD & Facility &  Note & Reference \\  \hline
1 & J011311.82+013542.4 & 0.2375 & 57282 & SDSS  &  59519 & SOAR   & turn-on & 1 \\ 
2 & J075544.35+192336.3 & 0.1083 & 53315 & SDSS  & 59527 & SOAR  & turn-on & 1 \\ 
3 & J102425.19+373903.0 & 0.1001 & 52998 & SDSS & 59603 & P200 & turn-on & 2 \\ 
4 & J120459.00+153513.8 & 0.2209 & 53467 & SDSS  & 59603 & P200  & non-CL & 2 \\ 
5 & J134148.78+370047.1 & 0.1968 & 53858 & SDSS  & 59604 & Keck  & turn-on & 2 \\ 
6 & J215055.73$-$010654.1 & 0.0879 & 53172 & SDSS  & 59519 & P200  & non-CL & 1 \\ \hline
1 & J102152.36+464515.7 & 0.2040 & 52614 & SDSS  & 56769 & BOSS & turn-off & 3 \\ 
2 & J143455.32+572345.3 & 0.3403 & 52441 & SDSS  & 57426 & WHT  & turn-off  & 4 \\   \hline  

\multicolumn{8}{p{12cm}}{Notes. The P200 and WHT refer to the 200-inch Hale telescope at Palomar observatory and the William Herschel Telescope in La Palma, respectively.  Reference: (1) \citet{Navas22}, (2) \citet{Navas23b}, (3)  \citet{Macleod16}, (4) \citet{Macleod19}. } 

\end{tabular}
\end{center}
\end{table*}

Follow-up spectroscopic observations were initiated in February 2022 using several facilities, including the 2.4m Hiltner telescope at the MDM Observatory, located at Kitt Peak, Arizona, US. The MDM observations were performed using the VPH red grism alongside a 4$^{\prime\prime}$ slit at the outer position. This configuration provided spectral coverage between 4000--9000$\,$\AA\ with a spectral resolution of 370,  albeit the system was most sensitive between 4500--8000$\,$\AA. The (on-source) exposure times were set between 2700 and 3600 seconds, depending on the target brightness. Calibration frames, including bias frames, sky flats, and Ar/Xe arc lamps were acquired during each night. 

Further follow-up observations were carried out between June 2022 to August 2023 with the Gemini-North 8m telescope, located in Manua Kea, Hawaii, US. We employed the GMOS spectrograph with a configuration that consisted of a 2$^{\prime\prime}$ slit, a R400 dispersor, a CG455 blocking filter (to remove high-order spectrum),  and a two-pixel binning mode in both spatial and wavelength direction. This configuration provided a spectral coverage of 5000--9500$\,$\AA\ with a spectral resolution $R\sim$480.  The sub-sample targeted by Gemini included relatively faint objects, i.e., with $g$ band magnitude fainter than 19.0. Exposure time was set between 180 and 600s depending on the brightness. Baseline observations, including standard stars, and Ar/Xe arc lamps, were acquired asynchronously as queue observations.  

More spectra were obtained from the Large Sky Area Multi-Object Fiber Spectroscopic Telescope (LAMOST) Quasar Survey \citep{Ai16,Dong18,Yao19,Jin23}, based on data from the low resolution spectrograph (LRS) on the $\sim$4$\,$m-effective-aperture Guoshoujing Telescope, located at Xinglong Observatory, China. The LAMOST Quasar Survey DR 8 contains spectra from September 2012 to May 2020. We only focused on spectra within two years prior to 2018 when ZTF observation was initiated, in order to approximately match the state reflected by ZTF light curves. We successfully found 4 objects with recent LAMOST spectrum available.  Each LAMOST spectrum was combined from a blue (3700--5900\AA) and red (5700--9000\AA) channel. For some objects, the average flux of blue and red channel could be very different caused by defect in the channel connection. We manually adjusted the relative flux level between two channels for these objects based on the SDSS photometry \citep{Jin23}, and left the overlapping region empty if the spectral shape inside this range was found peculiar.  

As for the historical spectra, all of our candidates were originally classified based on SDSS spectra. We found two targets in our sample with multi-epoch SDSS spectra available but their spectral shape were very similar and didn't show type transition upon visual inspection.  Therefore,  we only kept the spectrum which was close in time to the new epoch for the sake of constraining the CL timescale. All of these spectroscopic observations are summarized in Table \ref{tab:observation}.

We measured the \OIIIb\ emission line flux and aligned it over different epochs assuming \OIIIb\ are not variable during the timescale 10$\sim$20 years.  This assumption is reasonable because the spatial extent of the narrow line regions is known over kilo-parsec scale \citep{Bennert02} thus the timescale for narrow lines to show variation is more than 10$^3$ years.  Specifically, we performed initial spectral decomposition using {\tt PyQSOFit} package (see \S\ref{sec:results}) to extract \OIIIb\ flux. We adopted SDSS spectra as the flux reference because multiple-epoch SDSS observations display consistently robust flux calibration \citep{Green22}. The spectra from other facilities (instrument/telescopes) were re-scaled whenever the luminosity of \OIII\ are different by more than 10\% relative to SDSS values. 

\begin{figure*}[hb]
    \centering
    \includegraphics[width=0.85\textwidth]{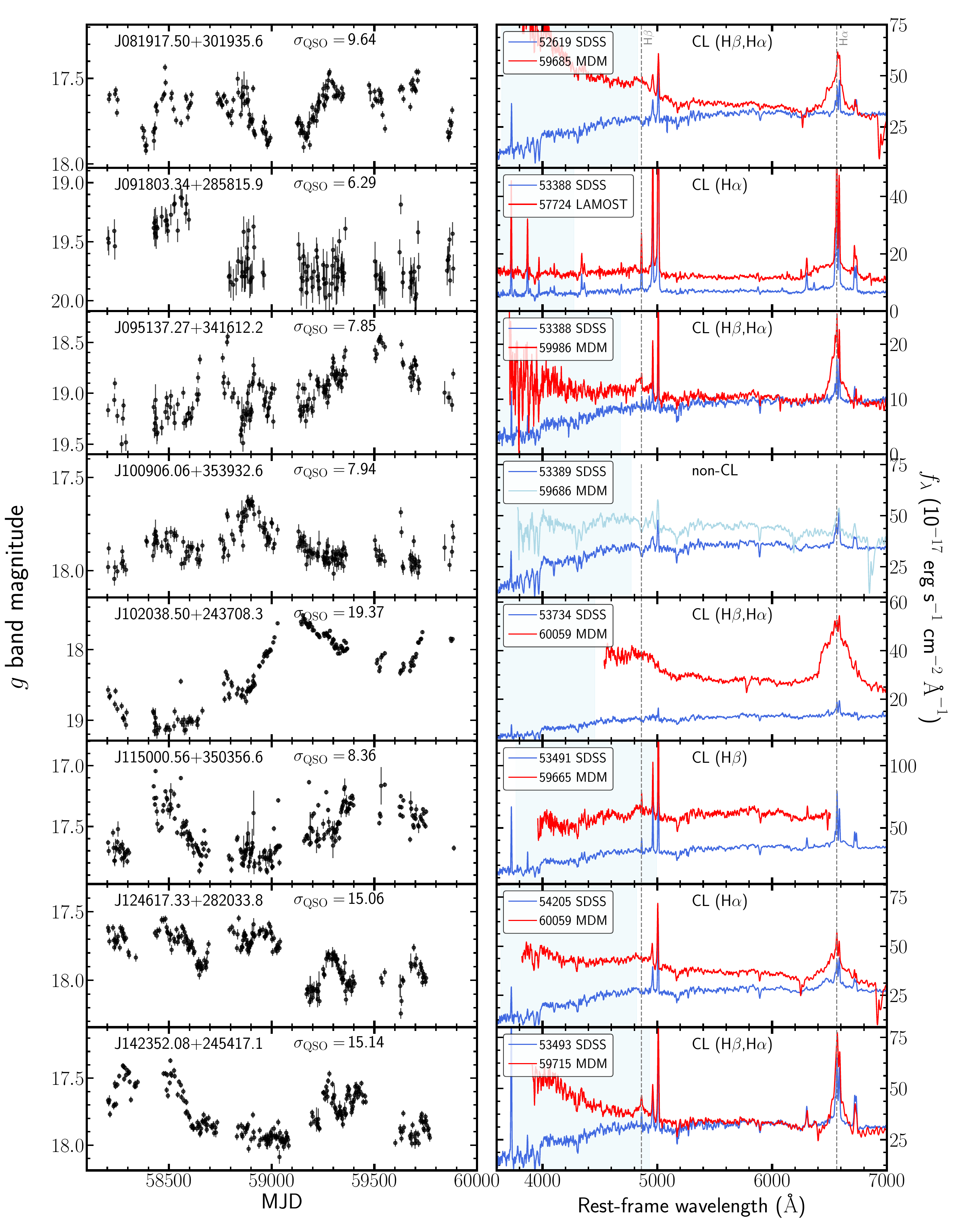}
    
    \caption{Spectroscopic follow-up results of 15 observed turn-on CL~AGN candidates. Left panels display the 3-day binned and cleaned ZTF $g$ band light curves, whose $\sigma_{\rm QSO}$ are labeled at the top. Right panels show the follow-up spectra compared with the historical spectra. We use red and blue (light blue) color to indicate bright and dim states, respectively.  Spectra from different instruments/telescopes are matched using \OIII\ flux.   The wavelength of \hbeta\ and \halpha\ are indicated using the vertical dashed lines. The blue shaded area indicates the range of the rest-frame $g$ band. As labeled at the top of right panels, 13 out of these 15 candidates are confirmed as turn-on CL AGNs while the rest two objects are disproved. }
    \label{fig:LC_Spectra_on}
\end{figure*}

\setcounter{figure}{2}
\begin{figure*}
    \centering
    \includegraphics[width=0.85\textwidth]{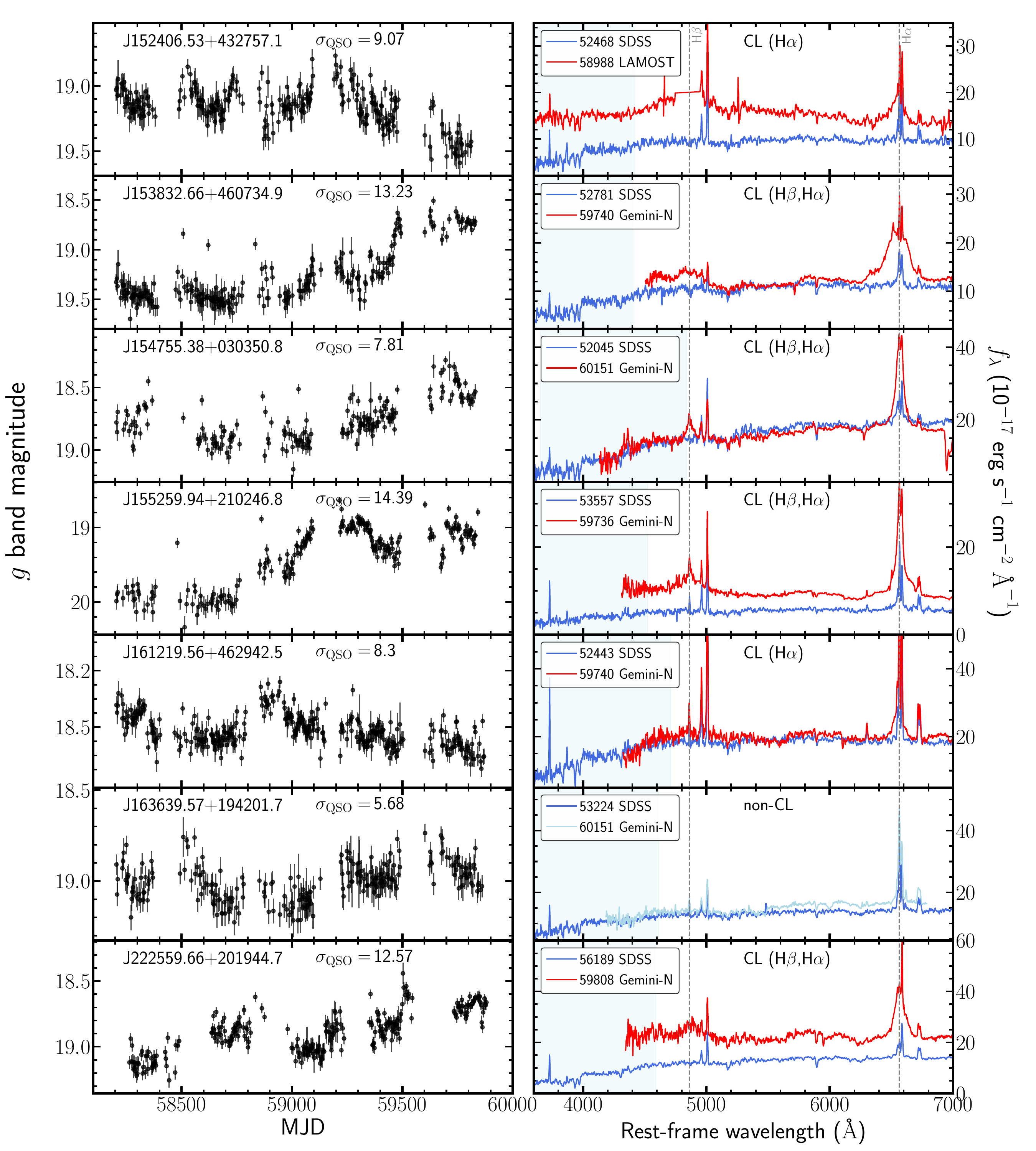}
    
    \caption{Continued}
\end{figure*}

\begin{figure*}[htbp]
    \centering
    \includegraphics[width=0.98\textwidth]{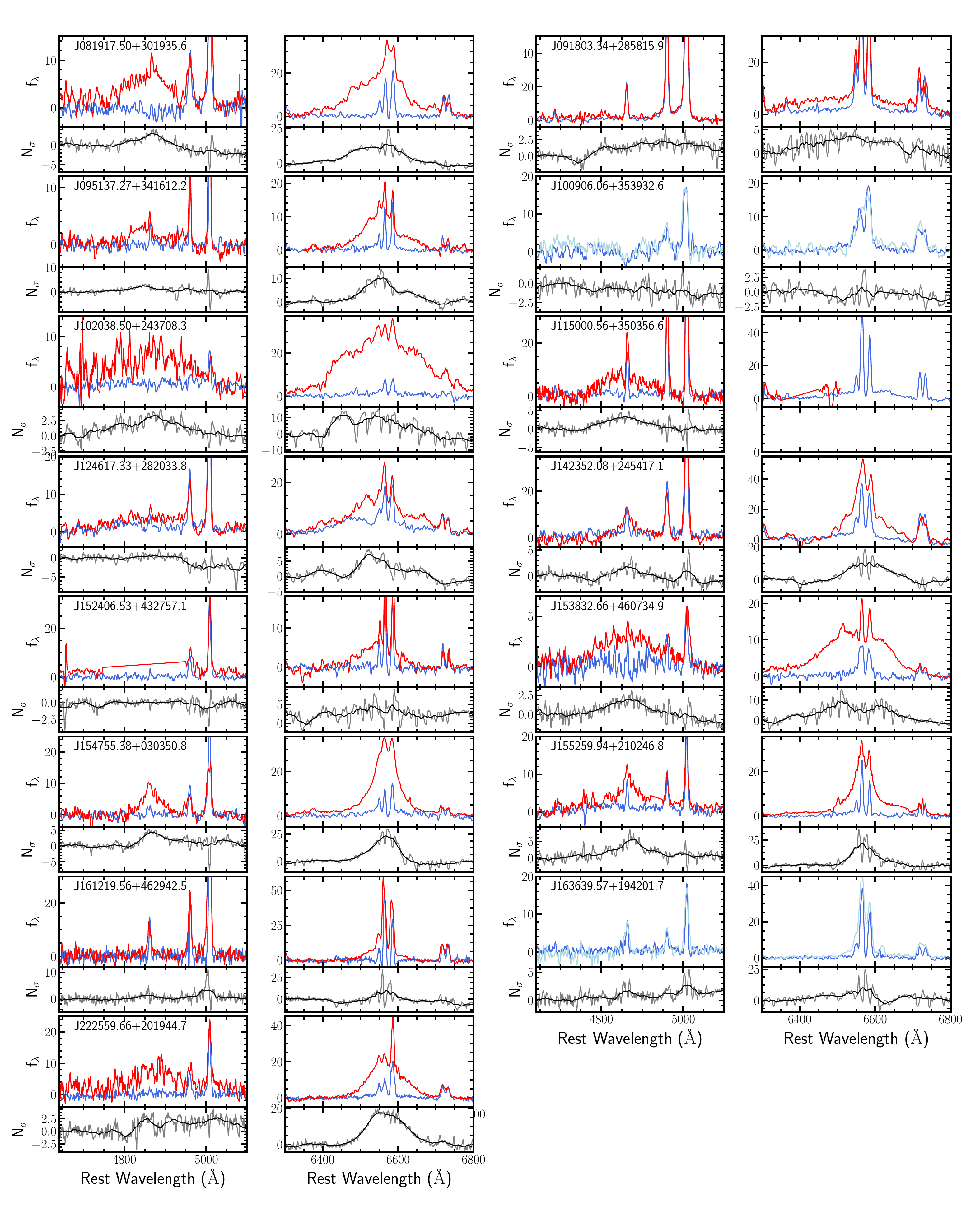}
 
    \caption{Comparison of broad-line profiles between bright (red) and dim (blue or light blue) state for 15 observed turn-on CL~AGNs candidates.  For each object, the left and right sub-panels represent the comparison of \hbeta\ and \halpha, respectively.  The bottom panels display the variability SNR $N_{\sigma}$ defined in Equation \ref{equ:Nsigma}, which represent the significance level of the variation. }\label{fig:profile_variation_on}
\end{figure*}

\begin{figure*}[htbp]
    \centering
    \includegraphics[width=0.82\textwidth]{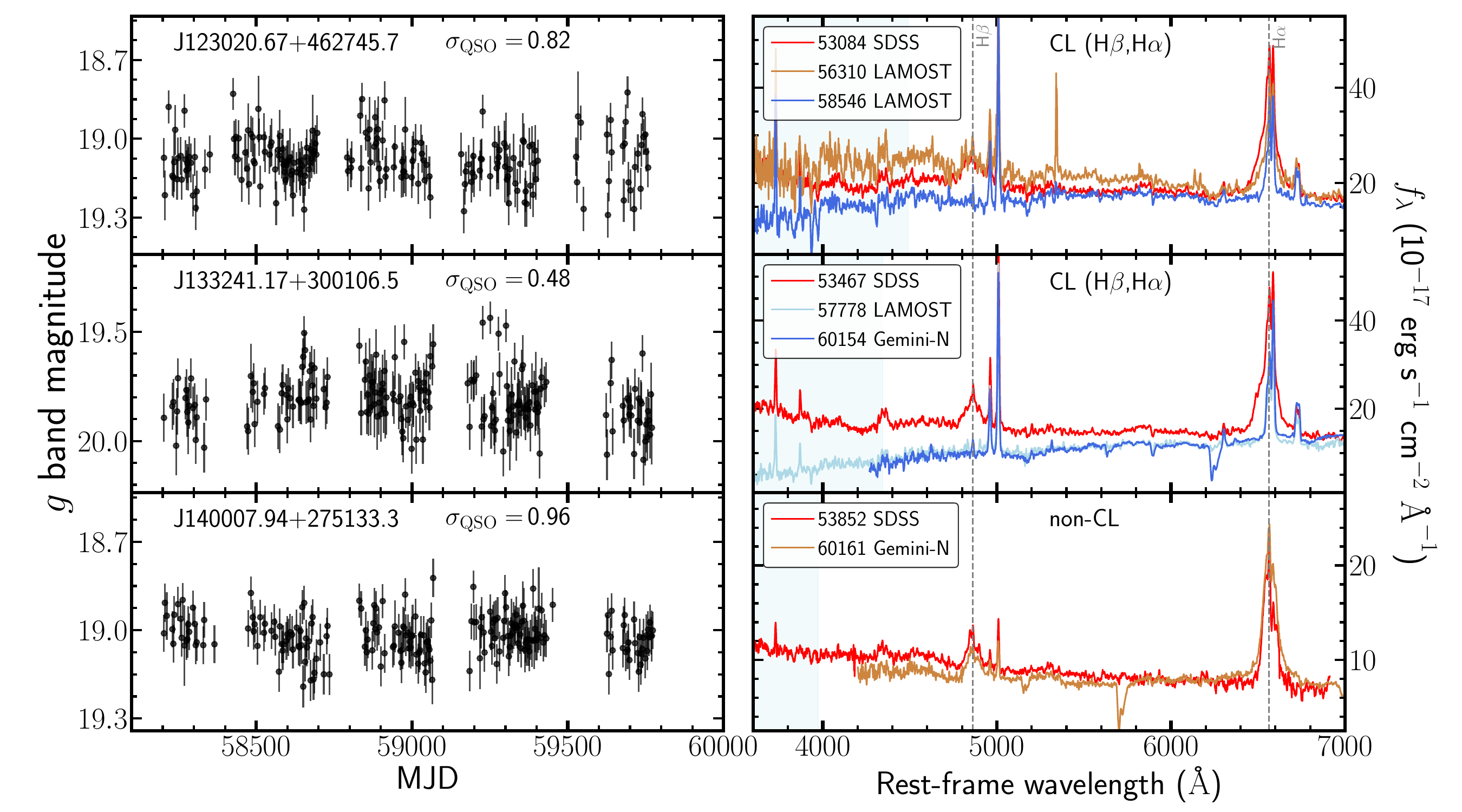}
    
    \caption{Spectroscopic follow-up results of 3 observed turn-off CL~AGN candidates. Left panels display the 3-day binned and cleaned  ZTF $g$ band  light curves. Right panels show the follow-up spectra and historical spectra. We use red (or brown) and blue (or light blue) to indicate bright and dim states, respectively.  Spectra from different instruments/telescopes are matched using \OIII\ flux.  The wavelength of \hbeta\ and \halpha\ are indicated using the vertical dashed lines.}
    \label{fig:LC_Spectra_off}
\end{figure*}

\begin{figure*}[htbp]
    \centering
    \includegraphics[width=0.98\textwidth]{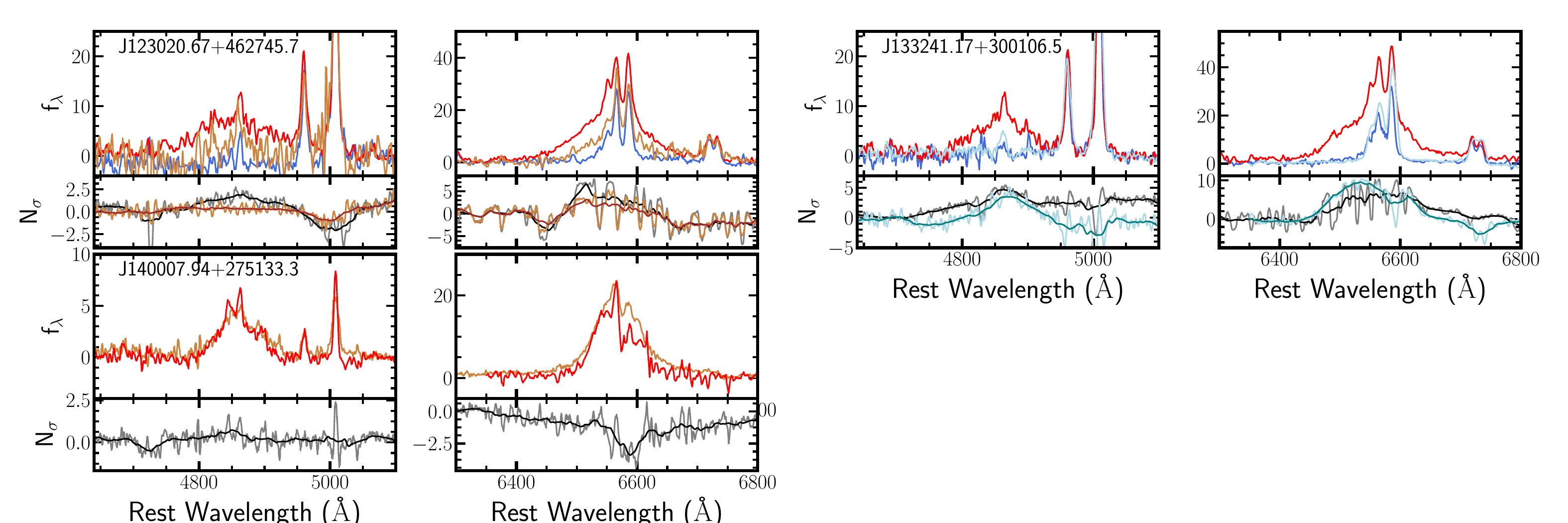}
    
    \caption{Same as Figure \ref{fig:profile_variation_on} but for the 3 observed turn-off CL~AGN candidates.}
    \label{fig:profile_variation_off}
\end{figure*}

\begin{table*}[h]
\caption{Properties of observed  CL AGNs candidates}\label{tab:CLAGNs}
\begin{center}
\footnotesize
\begin{tabular}{c c c  c  c   c  c    c  c   c c   c}
\hline \hline
  &    &  &   & &  \multicolumn{2}{c}{H$\beta$} &  \multicolumn{2}{c}{H$\alpha$}   & \\
   \cmidrule(r){6-7} 
   \cmidrule(r){8-9}
& Object name  &  z  & MJD &  State & $N_{\sigma}$  & log$L_{\rm H\beta}$   &  $N_{\sigma}$  & log$L_{\rm H\alpha}$   & log$M_{\rm BH}$ &  log$\frac{L_{\rm bol}}{L_{\rm Edd}}$  &   Notes \\  
&       &      &      &           &      &  (erg s$^{-1}$) &  & (erg s$^{-1}$)   & M$_{\odot}$ &  & \\ 

&    (1)   & (2) & (3)  & (4) & (5) & (6) & (7) &  (8) & (9)   & (10)  & (11)  \\ \hline
   
 \multicolumn{12}{c}{Turn-on CL AGNs} \\ \hline
1 & J081917.50+301935.6 & 0.0975 & 52619  &  D & \nodata  & $<40.1$ &   \nodata  & $<40.6$ & \nodata  & $<$$-2.87$  &   \nodata  \\ 
   &                  &        & 59685  & B & 3.03  & 41.37$\pm$0.03 &   13.66 & 41.94$\pm$0.01 & 7.87  & $-1.72$  &   \hbeta,\halpha\  \\ 
2 &  J091803.34+285815.9 & 0.2401 & 53388  & D & \nodata  & $<41.0$ &   \nodata  & 42.04$\pm$0.02 & \nodata  & $-2.41$  &   \nodata  \\ 
   &               &        & 57724   & B & 2.01 & $<41.5$ &   3.66  & 42.61$\pm$0.01 & 8.66 & $-1.92$  &   \halpha\ \\ 
3 & J095137.27+341612.2 & 0.1322 & 53388 & D & \nodata  & $<40.0$ &   \nodata  & $<40.5$ & \nodata  & $<$$-2.76$  &   \nodata  \\ 
   &                 &        & 59986   & B & 2.57  & 41.09$\pm$0.04 &   10.13  & 41.78$\pm$0.01 & 7.67  & $-1.65$  &  \hbeta(VIS), \halpha\ \\ 
4 & J100906.06+353932.6 & 0.1101 & 53389 & D & \nodata & $<40.2$ &   \nodata  & 40.76$\pm$0.07 & \nodata  & \nodata  &   \nodata  \\ 
   &                    &        & 59686  & D & $0.00$  & $<40.4$ &  0.76  & 41.18$\pm$0.03 & \nodata  & \nodata  &   non-CL  \\ 
5 & J102038.50+243708.3 & 0.1894 & 53734 & D & \nodata  & 41.24$\pm$0.09 &   \nodata  & 41.42$\pm$0.04 & \nodata  & $-3.44$  &   \nodata  \\ 
   &                    &        & 60059   & B & 3.39  & 42.00$\pm$0.02 &  11.26  & 42.86$\pm$0.01 & 9.15  & $-2.19$  &   \hbeta,\halpha\ \\ 
6 & J115000.56+350356.6 & 0.0611 & 53491 & D & \nodata  & 40.29$\pm$0.16 &   \nodata & 40.84$\pm$0.02 & \nodata  & $-2.73$  &   \nodata  \\ 
   &                      &        & 59665   & B & 3.34  & 40.93$\pm$0.03 &   \nodata  & \nodata & 7.93  & $-2.22$  &    \hbeta\ \\ 
7 & J124617.33+282033.8 & 0.0995 & 54205 & D & \nodata  & 40.98$\pm$0.09 &   \nodata  & 41.54$\pm$0.02 & \nodata  & $-2.54$  &   \nodata  \\ 
   &                       &        & 60059  & B & 0.64  & 41.15$\pm$0.02 &  7.18 & 41.84$\pm$0.01 & 8.35  & $-2.28$  &   \halpha\  \\ 
8 & J142352.08+245417.1 & 0.0744 & 53493 & D & \nodata  & 40.55$\pm$0.06 &   \nodata  & 41.01$\pm$0.04 & \nodata  & $-2.46$  &   \nodata  \\ 
   &                       &        & 59715  & B & 1.80  & 40.80$\pm$0.02 &   10.51  & 41.56$\pm$0.01 & 7.81  & $-1.98$  &  \hbeta(VIS), \halpha\  \\ 
9 & J152406.53+432757.1 & 0.1984 & 52468  & D & \nodata & $<40.8$ &   \nodata & $<41.3$ & \nodata  & $<$$-2.41$  &   \nodata  \\ 
   &                       &        & 58988   & B & \nodata  & \nodata &  4.90  & 41.87$\pm$0.02 & 8.02  & $-1.92$  &   \halpha\  \\ 
10 & J153832.66+460734.9 & 0.2026 & 52781  & D & \nodata  & $<40.8$ &   \nodata  & $<41.2$ & \nodata  & $<$$-3.24$  &   \nodata  \\ 
     &                     &        & 59740  & B & 2.04  & 41.70$\pm$0.01 &  10.65  & 42.48$\pm$0.01 & 8.76  & $-2.41$  &  \hbeta(VIS), \halpha,  \\ 
11 & J154755.38+030350.8 & 0.0947 & 52045 & D & \nodata & $<40.0$ &   \nodata  & 40.85$\pm$0.03 & \nodata & $-2.47$   &   \nodata  \\ 
     &                    &        & 60151  & B & 4.30 & 40.97$\pm$0.01 &  23.23  & 41.80$\pm$0.01 & 7.69  & $-1.65$  &  \hbeta, \halpha\  \\ 
12 & J155259.94+210246.8 & 0.1717 & 53557 & D & \nodata & 40.94$\pm$0.08 &   \nodata  & $<40.9$ & \nodata  & $<$$-2.56$  &   \nodata  \\ 
      &                  &        & 59736  & B & 5.49  & 42.00$\pm$0.01 &  22.05  & 42.23$\pm$0.01 & 7.82  & $-1.41$  &   \hbeta, \halpha\  \\ 
13 & J161219.56+462942.5 & 0.1254 & 52443 & D & \nodata  & $<40.7$ &   \nodata & $<41.2$ & \nodata  & $<$$-2.06$  &   \nodata  \\ 
     &                     &        & 59740 & B & 1.39  & 40.74$\pm$0.04 &  9.72 & 41.48$\pm$0.05 & 7.58  & $-1.82$  &   \halpha\  \\ 
14 & J163639.57+194201.7 & 0.1496 & 53224 & D & \nodata & $<40.7$ &   \nodata  & $<40.9$ & \nodata & \nodata   &   \nodata  \\ 
     &                     &        & 60151  & D & 1.71 & $<40.5$ &  10.20  & $<41.6$ & \nodata  &  \nodata  &   non-CL \\ 
15 & J222559.66+201944.7 & 0.1554 & 56189  & D & \nodata  & $<40.5$ &   \nodata & 41.16$\pm$0.05 & \nodata  & $-2.64$  &   \nodata  \\ 
     &                    &        & 59808 & B & 2.58 & 41.80$\pm$0.02 &   17.58  & 42.15$\pm$0.01 & 8.12 & $-1.78$   &  \hbeta(VIS), \halpha\ \\  \hline

 \multicolumn{12}{c}{Turn-off CL AGNs} \\ \hline
1 & J123020.67+462745.7 & 0.1803 & 53084 & B & \nodata  & 41.91$\pm$0.03 &   \nodata  & 42.45$\pm$0.01 & 8.31  & $-1.71$  &   \nodata  \\ 
   &                      &        & 56310  & B & 0.46 & 41.26$\pm$0.14 &   3.16 & 42.09$\pm$0.01 & 8.26 & $-1.97$  &   \halpha\  \\ 
   &                      &        & 58546 & D & 1.96 & $<40.3$ &   6.59  & 41.77$\pm$0.03 & \nodata  & $-2.27$  &   \hbeta, \halpha\  \\ 
2 & J133241.17+300106.5 & 0.2204 & 53467 & B & \nodata  & 42.05$\pm$0.02 &   \nodata & 42.70$\pm$0.01 & 8.50  & $-1.69$  &   \nodata  \\ 
   &                     &        & 57778 & D & 4.67  & 40.94$\pm$0.34 &  6.75 & 41.78$\pm$0.03 & \nodata  & $-2.48$  &   \hbeta, \halpha\  \\ 
   &                     &        & 60154 & D & 3.51  & $<40.9$ &   9.52  & 41.73$\pm$0.04 & \nodata  & $-2.53$  &   \hbeta, \halpha\  \\ 
3 &J140007.94+275133.3 & 0.3338 & 53852  & B & \nodata & 42.02$\pm$0.01 &   \nodata  & 42.65$\pm$0.04& 8.16  & $-1.39$   &   \nodata  \\ 
   &                    &        & 60161 & B & 0.74  & 42.09$\pm$0.02 &  0.00  & 42.90$\pm$0.02 & 8.28 & $-1.29$   &   non-CL \\ 
  \hline
 \multicolumn{12}{p{18.0cm}}{\footnotesize Note.  Column (1) The name of the object. Column (2): Redshift. Column (3): MJD of the spectra. 
 Column (4): The state of the spectra, where B and D represent 'Bright' and 'Dim', respectively.  'Bright' and 'Dim' states are used to represent whether there is prominent broad emissions.  Column (5): The maximum value of the variation S/N $N_{\sigma}$ of \hbeta\ (see \S \ref{sec:new-CLAGNs}). Column (6): The \hbeta\ luminosity. Column (7) and (8) are same as Column (5) and (6) but for \halpha. Column (9):  the BH mass ($M_{\rm BH}$) calculated based on the \halpha\ luminosity and line width of the bright state. The bright state of J115000.56+350356.6 doesn't cover \halpha\ therefore we utilized \hbeta.  Column (10): Eddington ratio ($\frac{L_{\rm bol}}{L_{\rm Edd}}$) estimated based on the bright-state $M_{\rm BH}$. Column (11):  summary of the lines exhibiting CL phenomenon, while a 'non-CL' means neither \hbeta\ nor \halpha\ undertook a CL event. The VIS in the brackets means it is additionally confirmed through visual inspection. Turn-on CL AGNs are shown at the top while turn-off CL AGNs are shown at the bottom.} 
\end{tabular}

\end{center}
\end{table*}

Last, the recent spectra of several AGNs in our sample were reported by literature before this paper was written. LN22 and LN23b selected 30 turn-on candidates based on light curve auto classifier ALeRCE \citep{Sanchez21b,Forster21}. Among the ten objects common to both our sample and LN23b's, four objects were observed independently  by both LN23b and our group, and remaining six objects were additionally observed by LN23b (see Table \ref{tab:turn-on-candidates}).  In  turn-off AGN candidates, two objects were confirmed by \citet{Macleod16} and \citet{Macleod19}.  We summarize these additional observations  in Table \ref{tab:additional_observation}.  These objects are only involved in the calculation of success rates and the investigation of CL timescales (\S \ref{sec:CL_timescale}), but not in the Eddington ratio analysis due to inhomogeneous measurement.

\section{Results of pilot follow-up observations}\label{sec:results} \label{sec:new-CLAGNs}

We present the follow-up spectra of 15 turn-on candidates, compared with the historical spectra and ZTF $g$ band light curves in Figure \ref{fig:LC_Spectra_on}. We discover 13 out of 15 objects showing emergence of broad \hbeta\ and/or \halpha\ in the follow-up spectra, which is consistent with their current Type 1 like light curves. To make it clearer, hereafter we define the epochs with and without prominent Balmer emission lines as bright and dim states, respectively.   

We perform spectral decomposition for both bright and dim state spectra of each object, using the python package {\tt PyQSOFit}\footnote{\url{https://github.com/legolason/PyQSOFit}} \citep{Guo18,Shen19}.  The details and an example of our decomposition are presented in Appendix \ref{sec:decomposition}. Based on the decomposition result, we extract the emission-line profile by subtracting the best-fit AGN continuum model and the host galaxy component. Figure \ref{fig:profile_variation_on} compares the profiles between bright and dim states, for \hbeta\ and \halpha, respectively.  

To quantify the variation, we define a variation SNR $N_{\rm \sigma}$ following previous works \citep{Macleod19, Green22}: 

 \begin{equation}
     N_{\sigma}(\lambda) = (f_{\rm bright} - f_{\rm dim}) / \sqrt{\sigma_{\rm bright}^2 + \sigma_{\rm dim}^2} \label{equ:Nsigma}
 \end{equation}
where $f$ and $\sigma$ are the flux density and their uncertainty at $\lambda$, respectively. Then the $N_{\sigma}(\lambda)$ is smoothed by a running median with 32 \AA\ width in the rest-frame.  \citet{Macleod19} adopted $N_{\rm \sigma}({\rm H\beta})=3$ to distinguish CL and non-CL AGNs. In our sample, we find 12 (5) objects displaying $N_{\rm \sigma}\geq3$ for \halpha\ (\hbeta).  The $N_{\rm \sigma}$ measurements as well as the other spectral or AGN properties are presented in Table \ref{tab:CLAGNs}. 

This $N_{\sigma}$-based selection criteria is very sensitive to the S/N ratio of the spectrum. For instance, as shown in Figure \ref{fig:profile_variation_on}, J095137.27$+$341612.2 displays a clear emergence of a broad \hbeta\ component in the new spectrum. However, it is not identified as a CL AGN using $N_{\sigma}$-based definition due to not sufficient S/N. We thus supplement the $N_{\sigma}$-based selection with visual inspection,  by requiring a clear appearance of the \hbeta\ broad-line profile in the bright states. Additional 4 \hbeta\ turn-on CL AGNs are identified through our visual inspection. They are labeled as 'VIS' in the 'Notes' column in Table \ref{tab:CLAGNs} for clarification. 

In addition to the 13 turn-on CL AGNs confirmed based on our follow-up observation, another 4 candidates have been identified as turn-on CL AGNs by LN22 and LN23b, based on significant increase of \halpha\ equivalent widths or \halpha/\SII\ ratios (see LN23b for more details). Combining these data with our observation, in total we confirm 17 turn-on CL AGNs out of 21 candidates, suggesting a success rate of 81\% (see Table \ref{tab:Successrate-on}).

In Figure \ref{fig:LC_Spectra_off}, we present the follow-up spectra of 3 turn-off candidates, compared with the historical spectra and ZTF $g$ band light curves. The comparison of line profiles between bright and dim states are shown in Figure \ref{fig:profile_variation_off}, for \hbeta\ and \halpha, respectively. We combine the $N_{\sigma}$-based criteria and visual inspection to identify turn-off CL AGNs.
We discover two objects showing clear disappearance of their broad \hbeta\ and/or \halpha\ component, as consistent with their (almost) flat and scattered ZTF light curves. In addition, two more candidates have been identified by \cite{Macleod19} with a significant decrease of the \hbeta\ line profile with $N_{\sigma}>3$. Combining these additional data with our observation, we confirm a total of 4 turn-off CL AGNs from the 5 candidates, showing a success rate of 80\% (see Table \ref{tab:Successrate-on}).

Last, we discuss the four targets that we do not confirm as CL-AGNs. Based on our follow-up spectra, we find no evidence of broad emission lines for two targets, and the other two targets did not show broad-emission lines in the previous study by LN22 and LN23b (see Table \ref{tab:additional_observation}). J215955.73$-$010654.1 (from LN22) has recently transitioned back to a dim state according to the ZTF light curves, which may explain its current status; for  J163639.57$+$194201.7, it is possible that a broad component is present in the follow-up spectrum but the line strength is too low to definitely confirm; 
the remaining two candidates (J100906.06$+$353932.6 and J120459.00$+$153513.8) are Type 1.9 AGNs according to their SDSS spectrum. These Type 1.9 AGNs may still be capable of producing some variability in their light curves if the AGN continuum is not fully diluted by the flux of host galaxy. A noteworthy point about J100906.06+353932.6 is that the spectral shape of the follow-up spectrum indicates a stronger blue continuum than before, yet there is no increase in the broad \halpha\ emission line strength. 
A more rigorous selection for bona fide Type 2 AGNs and further requirement on the recent variability may increase the success rate.
 
In the case of turn-off candidates, one object (J140007.94$+$275133.3)  turns out to be non-CL AGN. Its low $\sigma_{\rm QSO}$ value may be attributed to low intrinsic variability. Avoiding such contamination is a challenge in variability-based selection, which might be a limitation of our turn-off selection methodology. 
Another potential problem is that light curves from the ZTF DR can hinder the AGN variations for low-$z$ very extended sources \citep{Sanchez23}.
Considering the small sample size of our pilot study, the actual success rate of turn-off selection process is more uncertain than the turn-on case, and could be potentially lower than what is obtained in this pilot survey.

\begin{table}[htbp]
\caption{Success rate}\label{tab:Successrate-on}
\begin{center}
\begin{tabular}{c c c c c}
\hline \hline
Type & candidates & confirmed & success rate \\ \hline
turn-on & 21 & 17 & 81\% \\
turn-off & 5 & 4 & 80\% \\
  \hline
\end{tabular}
\end{center}
\end{table}%

\section{Discussion}\label{sec:discussion}

\subsection{Comparison with other selection methods}

Our selection method based on the variability pattern analysis provides a high success rate of $\sim$80\%, which is more efficient than other methods. For example, the success rate of EVAs is 15$\sim$50\%, which is based on the continuum flux changes by a factor of 1.5$\sim$6
\citep{Macleod19}.   

This difference of the two methods is highlighted in Figure \ref{fig:LC_Spectra_on}, which shows clear appearance of broad emission lines but no significant flux changes in the optical continuum (e.g., J154755.38+030350.8). Some of the confirmed turn-on AGNs do not even display a noticeable blue continuum in their bright states (e.g., J154744.38$+$030350.8 and J161219.56$+$462942.5), possibly due to the large contrast between the host galaxy and AGN flux ratio. Such CL AGNs can be missed, if the selection method is based on continuum variability or color variation, while these objects are crucial for studying the CL mechanisms as they are likely to be close to the occurrence of the type transition.

Our success rate is also higher than that of the variability selection based on the light curve classifier, i.e., LN22 and LN23b, who selected a sample of 30 turn-on AGN candidates based on large variability detection by alert brokers and new classification as Type 1 AGNs by light curve classifier ALeRCE. They confirmed 50\% of their candidates as turn-on CL AGNs from follow-up observations. By examining the non-CL AGNs listed by LN23b, we find that these objects generally display low $\sigma_{\rm QSO}$ values and scattered ZTF DR light curves. As discussed by LN23b, some of them are likely false detections from ZTF alert stream due to bad image subtraction and/or mis-classified by ALeRCE, while other objects may be related to transient events, e.g., supernovae or tidal disruption events, which have returned to its dim state when the follow-up spectrum is taken.  On the other hand, we investigated the candidates missed by our selection but selected and confirmed by LN23b, finding that  their light curves indeed resemble Type 1 AGN light curves, while their $\sigma_{\rm QSO}$ are not sufficient to meet our criteria. This suggests that our selection method is based on more pronounced Type-1-like variability with a higher thresholds than those adopted by ALeRCE.

\subsection{Implication on the physical mechanism}

While the physical mechanism of CL AGNs is not yet fully understood, the most important constraints come from two observational signatures, i.e., the CL time scales and the characteristic Eddington ratio. In this section we discuss these two constraints in the context of the CL mechanism.

\subsubsection{CL~timescales}\label{sec:CL_timescale}

\begin{figure}[htbp]
    \centering
    \includegraphics[width=0.49\textwidth]{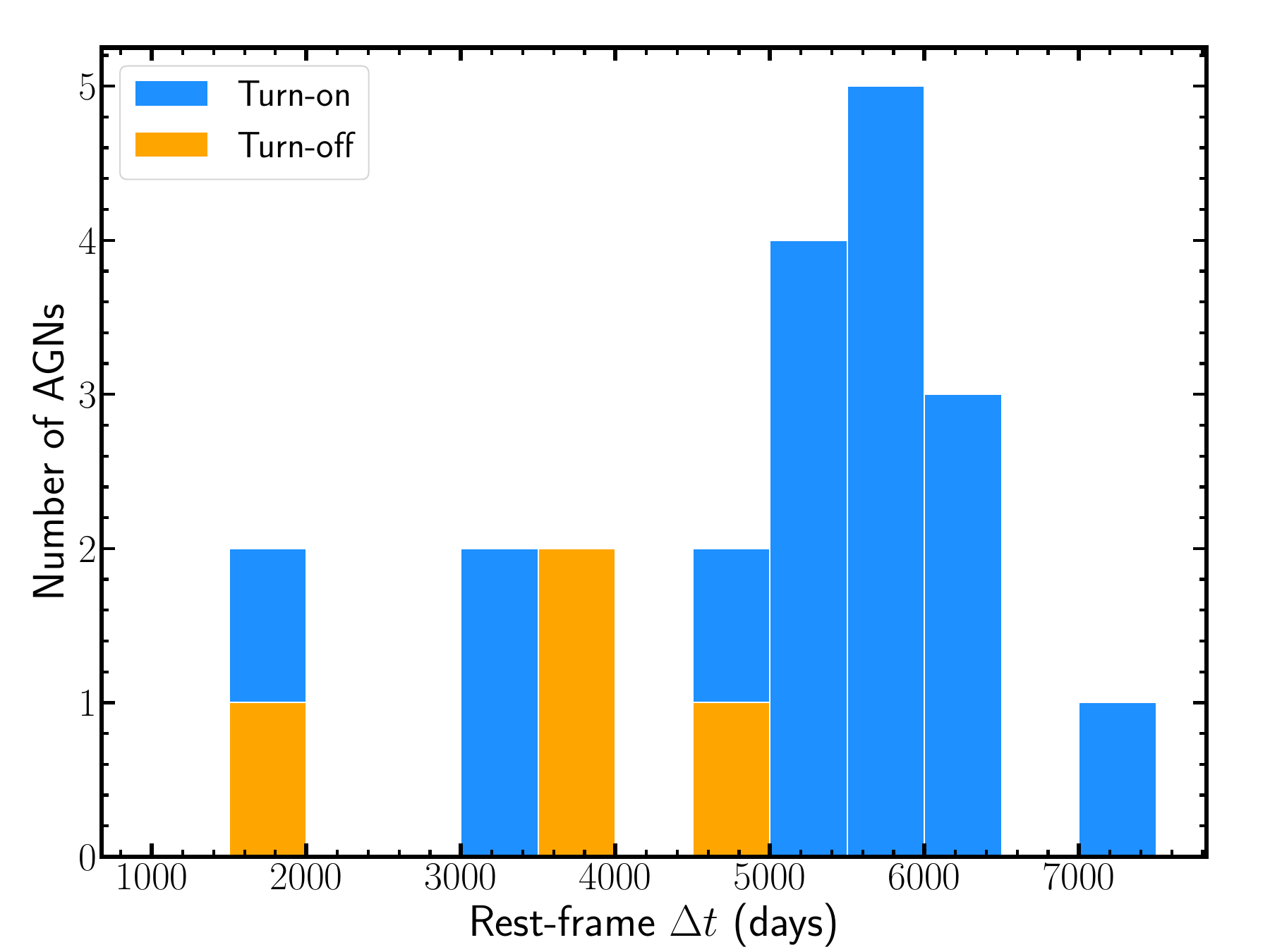}
    \caption{Distribution of the rest-frame time difference $\Delta t$ between the dim and bright state of confirmed turn-on (blue) and turn-off (orange) CL~AGNs. This time difference can be considered as the upper limit of CL timescale.  }
    \label{fig:time_difference}
\end{figure}

The CL timescale, which refers to the time required for broad emission lines to emerge/disappear, is typically 1$\sim$10 years \citep[see][and reference therein]{Ricci22}, and can be as short as several months in some extreme events \citep{Trakhtenbrot19a,Zeltyn22}. It's important to note that most of these estimates are based on the time difference between the dim and bright state spectra, therefore it should be considered only as the upper limits of the transition timescale.
Figure \ref{fig:time_difference} displays the distribution of the rest-frame time difference  $\Delta t$ of our sample, ranging from $\sim$1900 to $\sim$7400 days, corresponding to 5.2 to 20.2 years. These upper limits of CL timescale are generally consistent with previous works \citep{Yang18,Macleod19,Guo23Arxiv}.

These constraints are highly dependent on the timing of the spectroscopic observations, which are somewhat arbitrary. Thus, they are not a good tracer of the CL time scales. Accurate determination of transition time scale requires intensive spectroscopic monitoring  \citep[e.g.,][]{Green22,Zeltyn22} which is challenge to achieve. Based on our analysis, we conclude that the long and well-sampled photometric light curves can offer a valuable constraint for the  CL timescale, given that the change in variability pattern is a good tracer for CL events. 

\begin{figure}[htbp]
    \centering
    \includegraphics[width=0.49\textwidth]{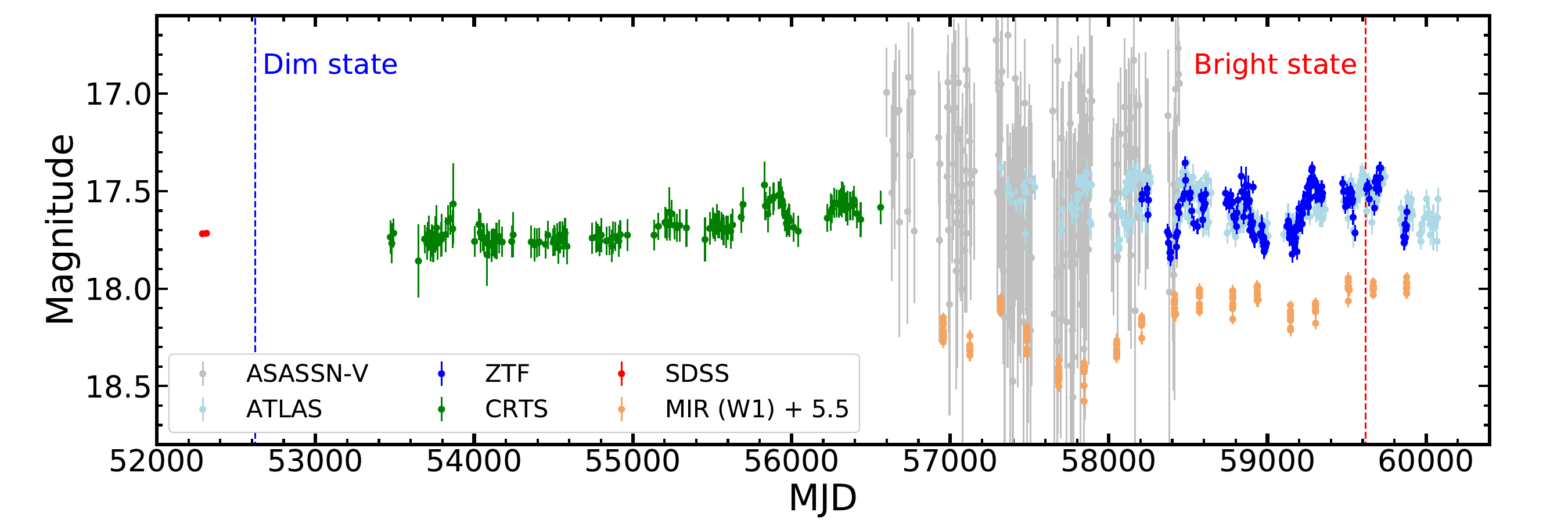}
    \includegraphics[width=0.49\textwidth]{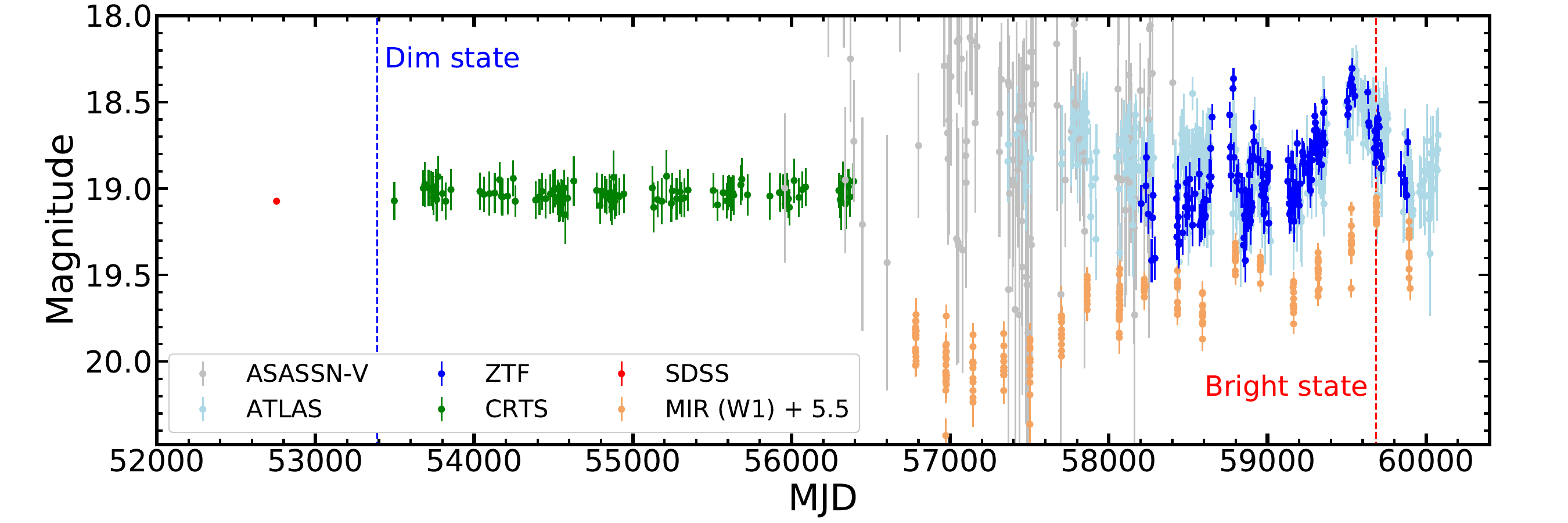}
    \includegraphics[width=0.49\textwidth]{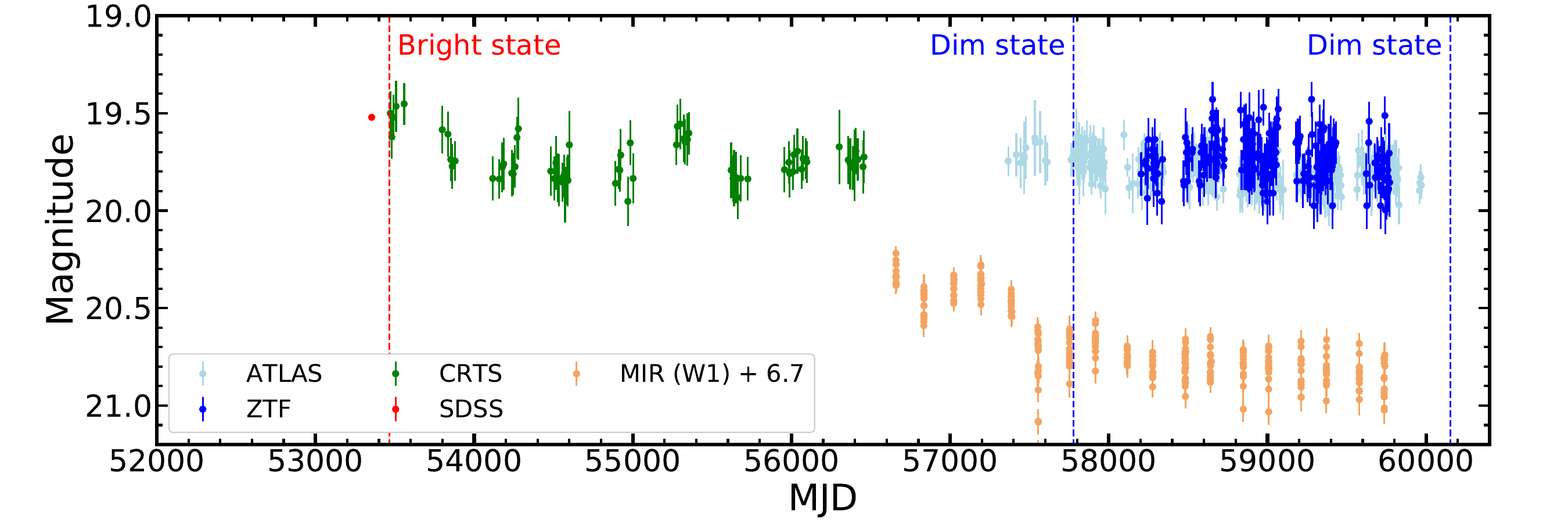}
    
    \caption{Three examples of the $\sim$20-yr optical light curves merged from different time domain surveys as presented in different colors. MIR W1 band light curves are also presented separately with a shift in magnitude for displaying purpose. Vertical dashed lines are used to indicate the spectroscopic observations, with blue and red color representing bright and dim states, respectively. From top to bottom, the target is J081917.50+301935.6, J095137.27+341612.2, and J133241.17+300106.5, respectively. }
    \label{fig:merged_LC}
\end{figure}

In Figure \ref{fig:merged_LC}, we combine $\sim$20-yr optical light curves from different time domain surveys, including the Catalina Real-Time Transient Survey  \citep[CRTS;][]{Drake09}, the All-Sky Automated Survey for Supernovae \citep[ASASSN;][]{Kochanek17}, the Asteroid Terrestrial-Impact Last Alert System  \citep[ATLAS;][]{Tonry18}, and ZTF.  To merge the light curves, we utilize the intercalibration software {\tt PyCALI} \citep{Li14} or manually match the average magnitude between closest seasons if there is no overlapp. Mid-infrared (MIR) light curves based on WISE W1 band are also presented for comparison.

The first exemplary object in Figure \ref{fig:merged_LC} is a turn-on CL AGN, which displays Type-1-like variability (log $(1+\sigma_{\rm QSO;CRTS})\simeq0.8$) from at least MJD$\simeq55000$ and perhaps even at/before MJD$\simeq53800$. It demonstrates that the actual timescale for such a type transition can be much smaller than the upper limit reported in the literature. The second object exhibits a flat light curves (log $(1+\sigma_{\rm QSO;CRTS})=0.29$) at least until MJD$\sim$56400, which is followed by a rapid increase in the MIR light curves at around MJD=57600. For this target, the inferred CL timescale could be as short as 1$\,$200 days. The third object exhibits Type 1 variability in both CRTS and W1 light curves until a decrease at around MJD=57500. It is then spectroscopically confirmed as a turn-off CL AGN at MJD=57778, suggesting that the CL timescale might be as small as 400 days.  These exploration demonstrates the potential to constrain the timing and  timescale of CL event from the photometric light curves based on variability patterns. However, it should be aware that the actual timing of the type transition still needs to rely on spectroscopic monitoring, while the photometric light curves provide an excellent indication for the timing of spectroscopy.

As discussed in the literature, the observed timescale is significantly shorter compared with the viscous timescale that required for major changes of accretion rate in standard thin discs ($\sim 10^4$ yrs \citep{Krolik99}).  \citet{Sniegowska20} shows that the radiation pressure instability in a narrow zone between the inner ADAF and outer standard discs can result in a significantly shorter timescale. Magnetically elevated accretion disks \citep{Dexter19}, as well as the combination of large-scale magnetic field with radiation pressure instability \citep{Pan21} are also proposed as effective means to reduce the timescale. In addition, the involvement of disk outflow can aid in removing angular momentum and further reduce the timescale \citep{Feng21,Wu23,Wada23}. As presented in Figure \ref{fig:merged_LC}, these long and good quality photometric light curves can provide a good opportunity to investigate these scenarios, especially with the advent of upcoming time domain surveys.

\subsubsection{Eddington ratios} \label{sec:Eddington}

Previous studies have found that CL AGNs preferentially occur at low Eddington ratios  \citep[e,g., $
\sim$ 1 percent;][]{Macleod19, Green22, Lyu22, Temple23}, which is an important characteristic of CL AGNs.  This Eddington ratio can be analogous to that of the state transition of X-ray binaries \citep{Noda18,Ruan19,Ai20, Yang23}.  When the accretion rate decreases to a few percent of Eddington limit, the accretion discs can transition from a standard disc to an advection dominated accretion flow (ADAF), associated with large amount decrease of far UV and soft X-ray ionizing photons.

\begin{figure}[htbp]
    \centering
    \includegraphics[width=0.49\textwidth]{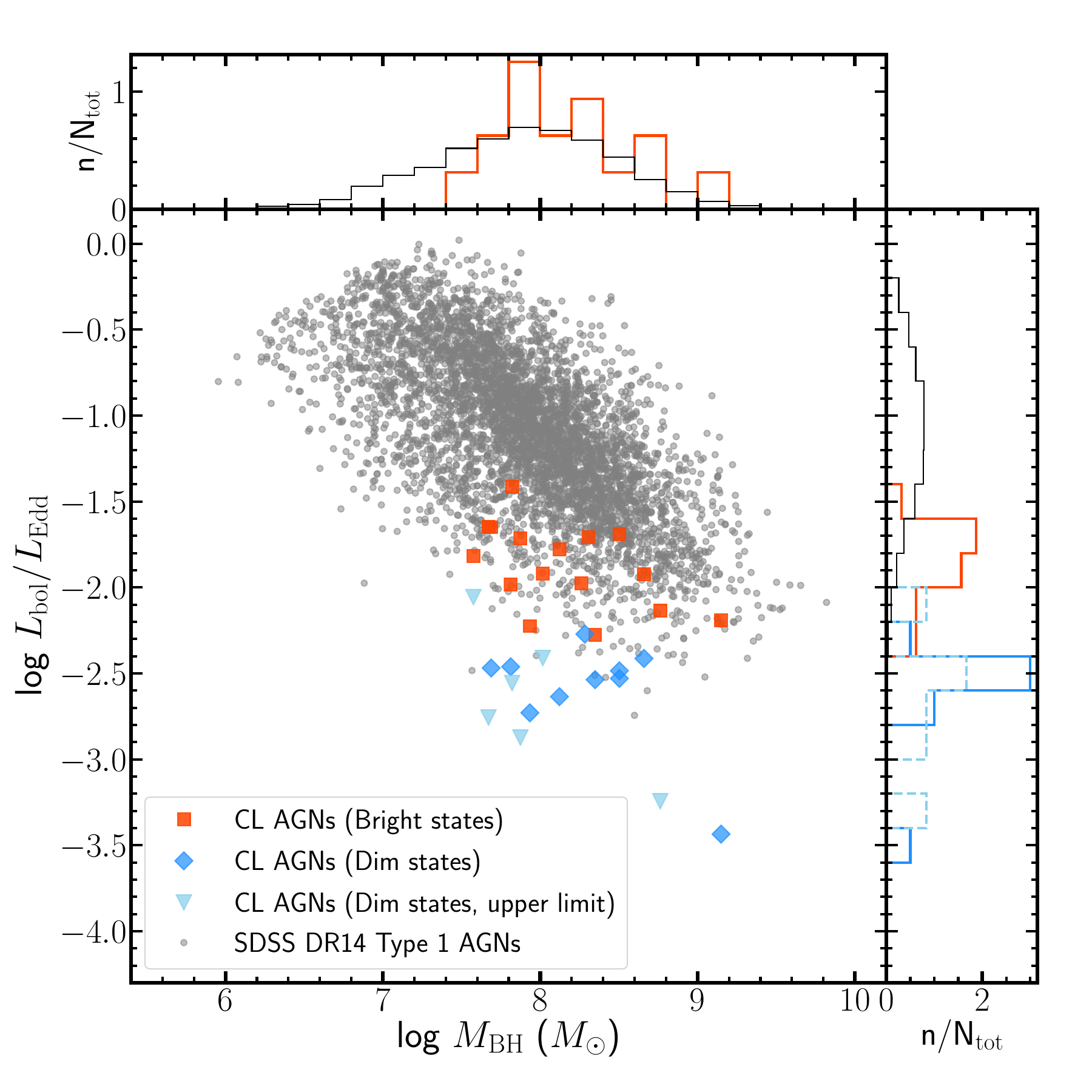}
    \caption{Distribution of our CL~AGNs on the $L_{\rm bol}/L_{\rm Edd}$--$M_{\rm BH}$ plane compared with Type 1 AGN control sample selected from SDSS DR14 Type 1 AGNs at $z<0.35$. The bright and dim states are displayed using red squares and blue diamonds. The light blue triangles indicate upper limits.  The upper and right panel represent the distribution of $M_{\rm BH}$ and Eddington ratio, respectively.}
    \label{fig:Edd_MBH}
\end{figure}

Figure \ref{fig:Edd_MBH} displays the distribution of our CL~AGNs on the $L_{\rm bol}/L_{\rm Edd}$--$M_{\rm BH}$ plane.  The BH mass is measured based on the H$\alpha$ luminosity and FWHM \citep{Reines13} from the bright states. The bolometric luminosity $L_{\rm bol}$ is calculated from $L_{5100}$ with a  bolometric correction factor of 9.26 \citep{Richards06}. The $L_{5100}$ is converted from $L_{\rm H\alpha}$ , based on the inverse of the Equation 2 from \citet{Reines13}:
\begin{equation}
L_{\rm 5100} =  \left( \frac{ L_{\rm H\alpha}} {5.52\times10^{42} {\rm \; erg\,s^{-1}}}  \right)^{1/1.157} \times 10^{44} {\rm \; erg\,s^{-1}}
\end{equation}

The bright states of our CL~AGNs exhibit a range of $L_{\rm bol}/L_{\rm Edd}$ from 0.005 to 0.04, while the dim states  show $L_{\rm bol}/L_{\rm Edd}$ of around 0.004 with some objects extending to lower than 0.001.    We construct a Type 1 AGN control sample from SDSS DR 14 \citep{Rakshit20} catalog by requiring redshift below 0.35, and  their $L_{\rm bol}/L_{\rm Edd}$ are measured based on the same method described above. We find that our CL AGNs are typically at the low $L_{\rm bol}/L_{\rm Edd}$ end of the Type 1 AGN distribution, which is consistent with previous findings \citep[e.g.,]{Macleod19}. Our results indicate that the transition should happen at $L_{\rm bol}/L_{\rm Edd}$ consistent or slightly lower than 0.01.

In the framework of disk-wind BLR formation scenario, the disappearance of broad emission lines at low Eddington ratio is an immediate consequence of the disappearance of high-latitude disk winds due to low column density at low $L_{\rm bol}/M_{\rm BH}^{2/3}$ \citep{Elitzur14}.  In our sample, we find that even for the dim states the bolometric luminosity is well above  5$\times10^{39}$ $(M_{\rm BH}/10^7M_{\odot})^{2/3}$ erg s$^{-1}$, which is the critical limit for appearance of BLRs, suggesting that this scenario is not playing a significant role in explaining these CL AGNs.   

\begin{figure}[htbp]
    \centering
    \includegraphics[width=0.49\textwidth]{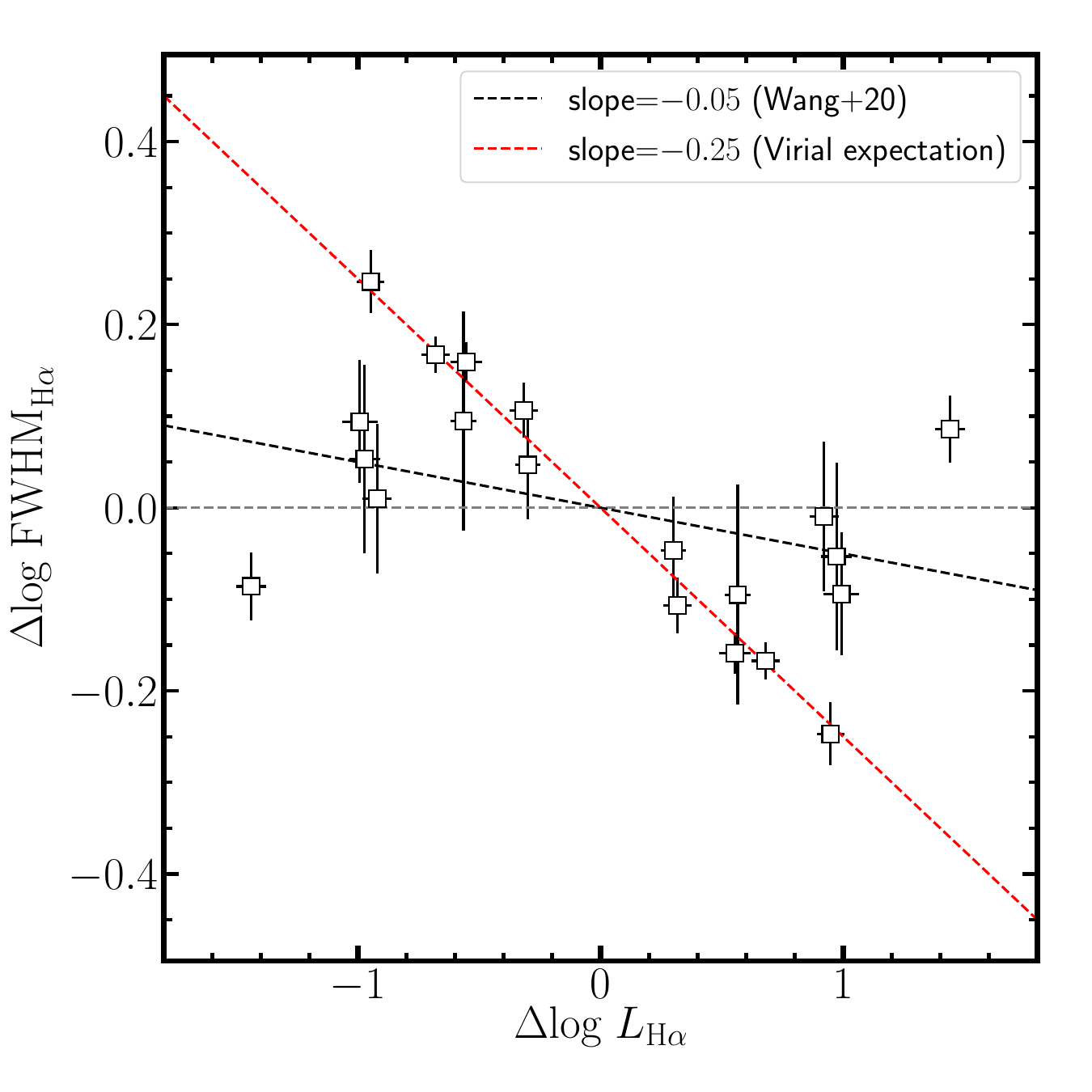}
    \caption{\halpha\ $\Delta {\rm FWHM}$--$\Delta L$ relation of CL AGNs based on objects with detectable \halpha\ in both dim and bright states. For each object, both the dim$-$bright and bright$-$dim results are presented. The red and black dashed line are the virial expectation and the relation derived for Type 1 AGNs from reverberation mapping studies \citep{Wang20}, respectively. The grey dashed line represents non-variation in line widths when luminosity changes.}
    \label{fig:breathing}
\end{figure}

In Figure \ref{fig:breathing}, we test the \halpha\ $\Delta {\rm FWHM}$--$\Delta L$ relation, i.e., how line widths respond to the changes in luminosity. For ideally virialized BLRs, the relation is supposed to display a slope of $-0.25$. In reverberation mapping studies, it is found that Balmer lines are generally consistent with the virial expectation  \citep[e.g.,][]{Barth15, Wang20, Fries23}. In Figure \ref{fig:breathing}, we present the result based on our confirmed CL AGNs with detectable \halpha\ in both dim and bright states. We plot both the dim$-$bright and bright$-$dim results for each object and make the figure symmetric around the origin. The majority of our targets show negative slopes and some of them closely follow the virial relation. Similar trends are also reported for other CL AGNs \citep{Green22, Zeltyn22, Neustadt23}. Such results may suggest that BLR can stay more or less virialized during the type transition, implying that CL AGNs are not likely caused by structural changes in BLRs.

\subsubsection{Blue excess}

We notice that several turn-on (off) CL~AGNs in our sample display asymmetric broad-line profile in their bright state spectra with flux excess in the blue side, e.g., J081917.50+301935.6,  J095137.27+341612.2, and J133241.17+300106.5.  These objects might resemble the CL AGN NGC 3516 \citep{Oknyansky21} which shows a similar strong blue asymmetry in Balmer lines.  Velocity-resolved reverberation mapping observations of NGC 3516 \citep{Oknyansky21,Feng21} reveal longer lags at blue side relative to red side, indicating the existence of inflows after turned-on. These two evidence collectively support a picture where a BLR with some bulk inflow existing, emits broad lines preferentially from the far side (thus blue side) of the BLR with respect to the observer. This preference on the far side can be resulted from the BLR self-obscuration effect \citep{Pancoast14,Oknyansky21}. 
Whether this picture of NGC 3516 can be applied to our sample is not clear, but the existence of a sample of CL AGNs with similar blue excess is interesting and requires future velocity-resolved reverberation mapping \citep[e.g.,][]{Feng21,Feng21b,Nagoshi23} to test their kinematic structures.

\subsubsection{Occurrence rate}

Given the success rate of our selection method, we estimate the occurrence rate of CL AGNs based on the distribution of $\sigma_{\rm qso}$ (Figure \ref{fig:illustration}). We identified 35 turn-on CL AGN candidates with log~(1+$\sigma_{\rm QSO})>0.8$ out of 9$\,$904 Type 2 AGNs, yielding a rate of 0.004 for CL AGN candidates. This rate can increase to 0.008 if we consider a relaxed threshold log~(1+$\sigma_{\rm QSO})>0.6$.
Then, we multiply the success rate of 81\% to the candidate rate, obtaining 0.003 to 0.007. For this simple estimate we assume a uniform success rate (81\%) across the rest of our selected candidates.
It suggests that roughly 0.003 to 0.007 of our Type 2 parent sample show significant Type-1-like variability. 
Note that the Type 2 parent sample may contain a small fraction of mis-classified weak Type 1 AGNs \citep[$\sim$5\%;][]{Navas23a}. The actual occurrence rate can be slightly higher than these estimates.
For turn-off CL~AGNs, the estimated occurrence rate is similar (i.e., $\sim$0.003). 

The estimated occurrence rate is broadly consistent with the previous works.  \citet{Yang18} reported an occurrence rate of 0.006\% (0.007\%) over a 0.9 to 12.6-year timescale by comparing SDSS multi-epoch spectra (SDSS-LAMOST two-epoch spectra). \citet{Green22} showed a rate of 0.03\% for CL occurring over $\sim$10 years.  LN22 and LN23b claimed a CL frequency of 0.3\% over a $\sim$15-year time span, which aligns with our result. They argued that this frequency can increase to 3\% considering that only objects with significant variability ($\sim$1/10) can trigger the alerts in their selection. \citet{Runco16} reported that 3\% of nearby Seyfert galaxies display clear disappearance of Balmer lines over $\sim$6 years, and 38\% of them show noticeable variation between intermediate types.  
\citet{Temple23} identified 21 known CL~AGNs in 749 hard X-ray selected AGNs, indicating a lower limit of CL occurrence rate of 1\% over 10$\sim$25-year timescales. \citet{Guo23Arxiv} estimated the \hbeta\ CL  occurrence rate to be 0.4\% over a timescale of 2$\sim$16 years by comparing SDSS and the Dark Energy Spectroscopic Instrument spectra \citep[DESI,][]{DESI23}.  The large difference between the previous studies suggests the existence of selection bias and difference in the CL definition.

These occurrence rates are estimated based on $\sigma_{\rm qso}$, which is derived from the DRW model. However, the DRW model may be overly simplistic to fully describe the quasar variability. For instance, many quasars show a significantly steep slope at high-frequency end of the power spectrum distribution \citep{Mushotzky11,Kasliwal15,Stone22}. The DRW process is the (1,0) order of the more general continuous time autoregressive moving average (CARMA) process \citep[e.g.,][]{Kelly14}. Higher-order CARMA processes, e.g., CARMA(2,1), have been suggested to better describe quasar light curves than DRW model \citep[e.g.,][]{Kasliwal17}. In addition, more complex models with inclusion of non-stationary processes are in principle better descriptions of these AGN light curves, e.g., the continuous time autoregressive fractionally integrated moving average \citep[CARFIMA;][]{Tsai2005}. More accurate estimations of the occurrence rate can be achieved with the development of a new methodology to distinguish the variability of Type 1 and 2 AGNs based on these more comprehensive models. While this is beyond the scope of this paper, it represents a promising direction for future research.

\section{Summary}\label{sec:summary}

In this work,  we performed a systematic variability study of a large sample of spectroscopically classified Type 1 and 2 AGNs, based on the ZTF light curves.  We demonstrated that these two classes can be well separated by the  variability metric $\sigma_{\rm QSO}$ (Figure \ref{fig:illustration}) which quantifies the resemblance of a light curve to a DRW model. However, a small subset of both Type 1 and Type 2 AGNs exhibit variability patterns inconsistent with their previous classification, suggesting the occurrence of CL events. Based on the criteria log~$(1+\sigma_{\rm QSO})>0.8$ and $<0.3$, we selected 35 turn-on and 12 turn-off CL AGNs, respectively.

We spectroscopically confirmed 17 (4) turn-on (turn-off) CL~AGNs out of 21 (5) candidates based on our follow-up observations using the Gemini-North 8m and MDM 2.4m telescope, or available spectra in the literature \citep[LN22,LN23b]{Macleod19}. The high success rate ($\sim$80\%) of our pilot observations demonstrates the advantage of our selection method based on the variability pattern analysis, compared to the previous selection method.
Our approach is also capable of identifying CL AGNs without noticeable appearance of the blue continuum, that is easily missed by the selection based on large variability amplitude or large change in colors. 

We found that the time difference between the dim and bright states ranges from 5.2 to 20.2 yrs (Figure \ref{fig:time_difference}), which provides an upper limit to the transition timescale. However, our investigation of the variability patterns based on $\sim$20-yr optical/MIR light curves (Figure \ref{fig:merged_LC}) suggests that the actual timescale can be considerably smaller than these upper limits. 
In addition, we confirmed that the typical transition $L_{\rm bol}/L_{\rm Edd}$ is approximately $\sim 0.01$ or below (Figure \ref{fig:Edd_MBH}). We found that the CL~AGNs occurrence rate is roughly 0.3 \% for $z<0.35$ AGNs over 5$\sim$20 years.

In this work,  we compared the historical spectroscopic classification with current variability information. Our exploration in \S \ref{sec:CL_timescale} demonstrates the great potential to identify changing-look AGNs based on the change of the variability patterns, given the consistency between the spectral type and the variability patterns. 
The well-sampled light curves over long time baseline from the future time domain surveys will be extremely effective to identify CL AGNs. For example, the Wide Field Survey Telescope \citep[WFST,][]{WFST23} and the Legacy Survey of Space and Time \citep[LSST,][]{LSST09, Ivezic19} will provide a crucial opportunity for studying CL AGNs. 
The LSST Wide-Fast-Deep (WFD) survey will observe a $\sim$18$\,$000 square degree sky area for $\sim$1$\,$000 times over 10 years,
providing the data for $\sim$ten million AGNs. Considering a broad range of the CL~AGN occurrence rate of $0.1\sim3$\% over 10 years,  we anticipate ten $\sim$ hundred thousands of CL~AGNs to be detected by the LSST.

\begin{acknowledgments}
We thank the anonymous referee for the helpful comments and suggestions.
We thank Nathaniel Butler, Junjie Jin, Yuming Fu, Huimei Wang for helpful discussion. This work is supported by the National Research Foundation of Korea (NRF) grant funded by the Korean government (MEST) (No. 2019R1A6A1A10073437), the Basic Science Research Program through the National Research Foundation of Korean Government (2021R1A2C3008486), and the National Key R\&D Program of China No.2022YFF0503402. This work is based on observations obtained at the Gemini North Observatory, which is operated by the Association of Universities for Research in Astronomy (AURA) under a cooperative agreement with the NSF on behalf of the Gemini partnership: the National Science Foundation (United States), the National Research Council (Canada), CONICYT (Chile), the Australian Research Council (Australia), Ministerio da Ciencia e Tecnologia (Brazil) and Ministerio de Ciencia, Tecnologia e Innovacion Productiva (Argentina). This work is based on observations obtained with the Samuel Oschin Telescope 48 inch and the 60 inch Telescope at the Palomar Observatory as part of the Zwicky Transient Facility project. ZTF is supported by the National Science Foundation under Grant No. AST-2034437 and a collaboration including Caltech, IPAC, the Weizmann Institute for Science, the Oskar Klein Center at Stockholm University, the University of Maryland, Deutsches Elektronen-Synchrotron and Humboldt University, the TANGO Consortium of Taiwan, the University of Wisconsin at Milwaukee, Trinity College Dublin, Lawrence Livermore National Laboratories, and IN2P3, France. Operations are conducted by COO, IPAC, and UW. This research has made use of the NASA/IPAC Infrared Science Archive, which is funded by the National Aeronautics and Space Administration and operated by the California Institute of Technology. This work makes use of data products obtained from the Guoshoujing Telescope (the Large Sky Area Multi-object Fiber Spectroscopic Telescope, LAMOST). LAMOST is a National Major Scientific Project built by the Chinese Academy of Sciences. Funding for the project has been provided by the National Development and Reform Commission. LAMOST is operated and managed by the National Astronomical Observatories, Chinese Academy of Sciences.
\end{acknowledgments}

\facility{IRSA, ZTF, Gemini, MDM, LAMOST} 

\software{Astropy \citep{Astropy13,Astropy18,Astropy22}, {\tt PyQSOFit} \citep{Guo18}, {\tt qso\_fit} \citep{Butler11}, {\tt PyCALI} \citep{Li14}.}

\appendix
\renewcommand\thefigure{\thesection.\arabic{figure}}  
\renewcommand\thetable{\thesection.\arabic{table}}  

\section{Description and justification of selected candidates}\label{sec:appendixA}
\setcounter{figure}{0}   
\setcounter{table}{0}   

Table \ref{tab:turn-on-candidates} and \ref{tab:turn-off-candidates} provides the properties of our selection turn-on and turn-off candidates, respectively. 

\begin{figure*}[htbp]
    \centering
    \includegraphics[width=0.45\textwidth]{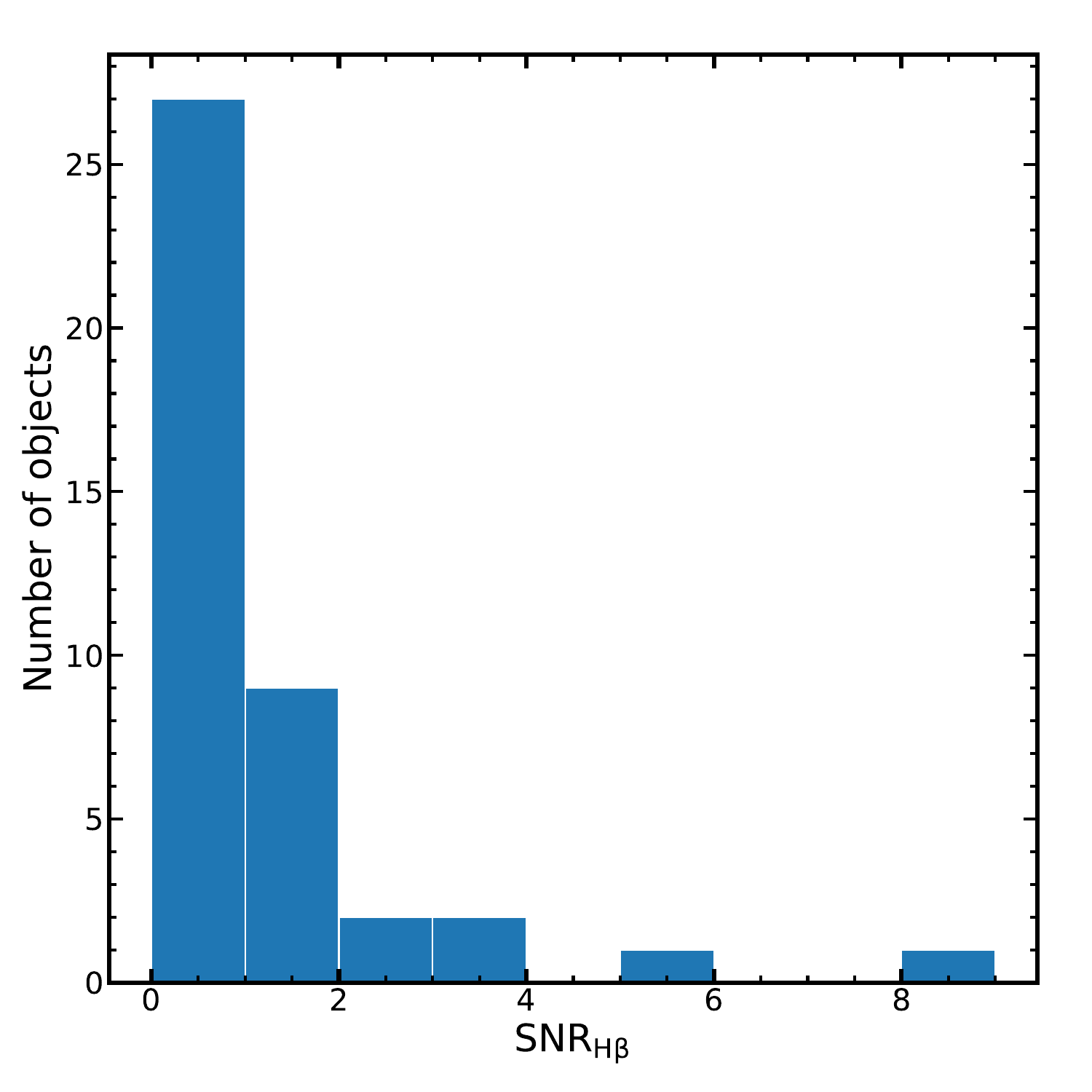}
    \includegraphics[width=0.45\textwidth]{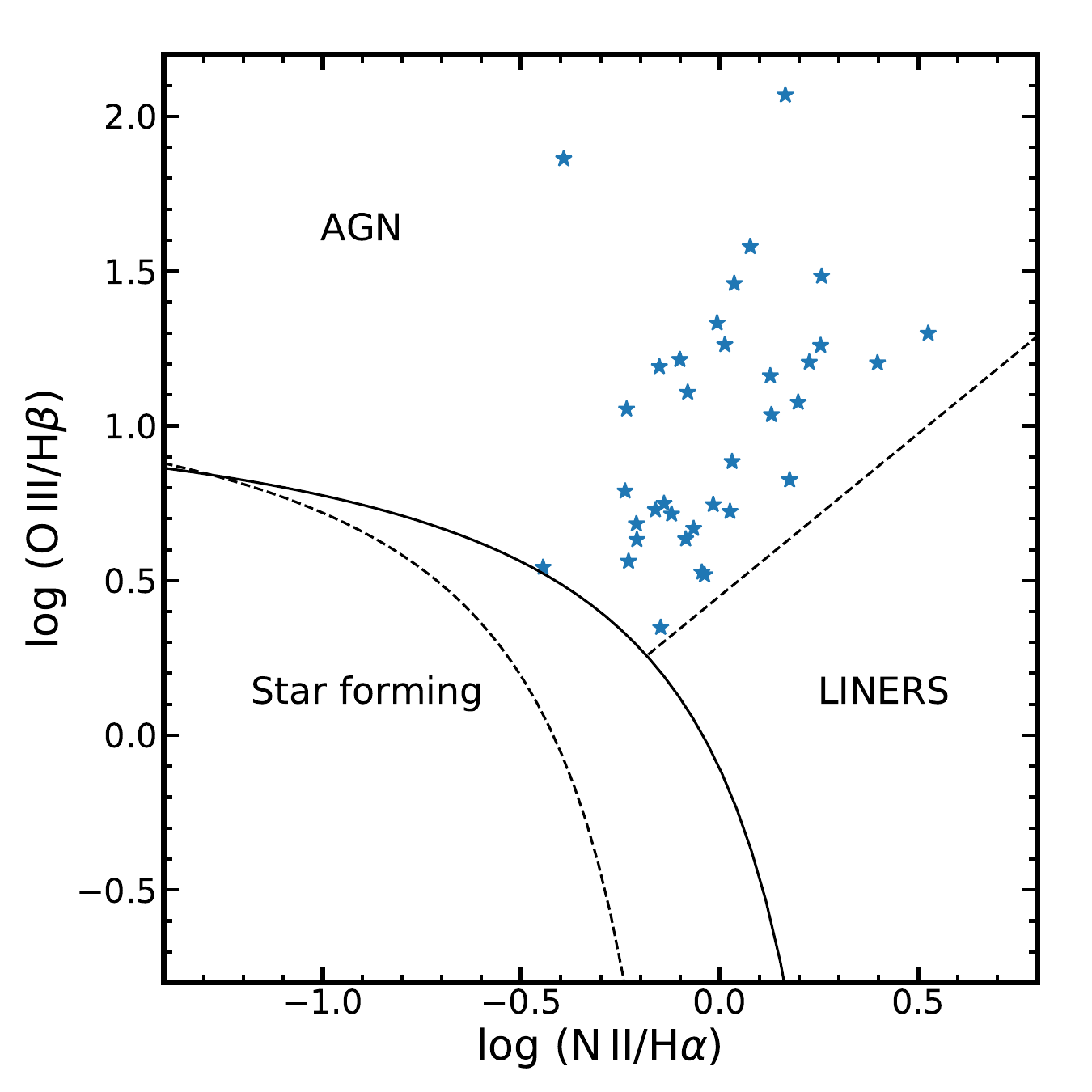}
    \caption{Left panel: SNR of broad \hbeta\  in the dim state for 42 Type 2 AGNs selected as turn-on candidates. We exclude objects with obvious broad \hbeta\ suggested by SNR$_{\rm H\beta}>2$. Right panel: BPT diagram for turn-on candidates.  This plot confirms the nature of our candidates to be Type 2 AGNs rather than star forming galaxies or LINERS.  }
    \label{fig:SNR_hbha}
\end{figure*}

Below we justify our selected turn-on candidates. First, we check our selected turn-on candidates whether there is a broad component. We define a SNR of broad \hbeta,  which is the ratio between the mean flux and the standard deviation within the a window with 1 FWHM width surrounding the peak of broad lines. Because most of the previous CL studies focus on \hbeta,  we exclude sources with SNR$_{\rm H\beta}$ larger than 2 in the following studies, based on which six objects are removed.   The distribution of SNR$_{\rm BH\beta}$ are shown in Figure \ref{fig:SNR_hbha}.  We checked the information of these mis-classified AGNs listed in MQC. We found that these mis-classification was inherited from SDSS DR 16 where they were labeled as narrow line galaxies/quasars by SDSS pipeline.

Second, because our turn-on CL candidates are selected from MQC, where a small portion of star forming galaxies as well as LINERS can contaminate our parent sample. Here we justify our selected turn-on candidates by checking the BPT diagnostics of these candidates \citep{Baldwin81}. As shown in the right panel of Figure \ref{fig:SNR_hbha},  we find that all of these candidates are Type 2 AGNs rather than star-forming galaxies or LINERS.

\begin{table*}[htp]
\caption{Turn-on CL~AGN candidates} \label{tab:turn-on-candidates}
\begin{center}
\begin{tabular}{ c c c c c c c c c c c}
\hline \hline
   SDSS identifier & RA & DEC & $z$ & $g_{\rm MED}$ & $\sigma_{\rm QSO}$ & $\sigma_{\rm NotQSO}$ & $\sigma_{\rm var}$  & Confirmed? & Reference  \\ 
                            & degree &  degree  &    &  mag &   &       &       &   &      \\  \hline 
J003741.36$+002906.8$ & 9.42234   & $0.48526$  & 0.1506 & 19.17 & 6.16 & 3.54 &  23.06 &  \nodata & \nodata \\ 
J011311.82$+013542.4$ & 18.29924  & $1.59516$  & 0.2375 & 19.27 & 16.30 & 1.45 &  50.94 &  Yes & LN22 \\ 
J021159.31$-001031.9$ & 32.99718  & $-0.17547$ & 0.2421 & 19.85 & 6.81 & 3.84 &  26.42 &  \nodata & \nodata \\ 
J075544.35$+192336.3$ & 118.93482 & $19.39344$ & 0.1083 & 18.31 & 6.75 & 8.91 &  50.21  &  Yes & LN22  \\ 
J081147.98$+382154.3$ & 122.94996 & $38.36510$ & 0.1873 & 19.18 & 5.86 & 2.09 &  17.10  &  \nodata & \nodata \\ 
J081917.50$+301935.6$ & 124.82294 & $30.32660$ & 0.0975 & 17.68 & 9.64 & 7.39 &  53.28  &  Yes & this work, LN23b \\ 
J091803.34$+285815.9$ & 139.51395 & $28.97118$ & 0.2401 & 19.70 & 6.29 & 1.52 &  19.13  &  Yes & this work \\ 
J095137.27$+341612.2$ & 147.90527 & $34.27011$ & 0.1322 & 19.00 & 7.85 & 8.63 &  47.18  &  Yes & this work \\ 
J100742.24$+333136.4$ & 151.92606 & $33.52679$ & 0.2573 & 19.36 & 5.47 & 1.12 &  13.55  &  \nodata & \nodata \\ 
J100906.06$+353932.6$ & 152.27534 & $35.65907$ & 0.1101 & 17.88 & 7.94 & 3.69 &  29.90  &  No & this work \\ 
J102038.50$+243708.3$ & 155.16042 & $24.61900$ & 0.1894 & 18.19 & 19.37 & 8.87 &  139.15  &  Yes & this work \\ 
J102425.19$+373903.0$ & 156.10500 & $37.65092$ & 0.1001 & 18.72 & 5.58 & 3.15 &  19.13  &  Yes & LN23b \\ 
J112634.19$+395539.7$ & 171.64249 & $39.92773$ & 0.1918 & 20.00 & 5.74 & 2.64 &  16.77  & \nodata& \nodata \\ 
J115000.56$+350356.6$ & 177.50233 & $35.06577$ & 0.0611 & 17.53 & 8.36 & 16.49 &  103.85  &  Yes   & this work \\
J120459.00$+153513.8$ & 181.24583 & $15.58719$ & 0.2209 & 19.60 & 6.05 & 3.74 &  22.83 &  No & LN23b \\ 
J122026.75$+363327.8$ & 185.11150 & $36.55777$ & 0.1003 & 18.68 & 8.22 & 4.52 &  34.08 &  \nodata& \nodata \\ 
J124617.33$+282033.9$ & 191.57228 & $28.34276$ & 0.0995 & 17.80 & 15.06 & 4.99 &  62.93  &  Yes & this work \\ 
J134148.78$+370047.1$ & 205.45328 & $37.01311$ & 0.1968 & 19.07 & 6.71 & 4.93 &  24.27  &  Yes & LN23b \\ 
J134154.56$+294058.5$ & 205.47737 & $29.68297$ & 0.0446 & 18.00 & 8.13 & 17.31 &  117.02  &  \nodata & \nodata \\ 
J135007.69$+124657.3$ & 207.53209 & $12.78262$ & 0.1412 & 19.36 & 5.99 & 3.41 &  22.42  &  \nodata & \nodata \\ 
J141415.17$+264451.4$ & 213.56326 & $26.74764$ & 0.0349 & 15.72 & 5.64 & 6.12 &  31.88  &  \nodata& \nodata \\ 
J142352.08$+245417.1$ & 215.96706 & $24.90476$ & 0.0744 & 17.84 & 15.14 & 8.82 & 96.02  &  Yes & this work \\ 
J144021.49$+141125.7$ & 220.08955 & $14.19051$ & 0.1220 & 18.65 & 5.64 & 8.93 &  40.16  &  \nodata & \nodata \\ 
J144344.18$+591040.5$ & 220.93410 & $59.17793$ & 0.1401 & 18.72 & 9.68 & 3.95 &  33.36  &  \nodata& \nodata \\ 
J152406.53$+432757.1$ & 231.02727 & $43.46591$ & 0.1984 &  19.14 & 9.07 & 0.94 &  19.35  &  Yes & this work \\ 
J153832.66$+460734.9$ & 234.63612 & $46.12639$ & 0.2026 & 19.41 & 13.23 & 4.56 &  56.72  &  Yes & this work, LN23b \\ 
J154755.38$+030350.8$ & 236.98079 & $3.06414$  & 0.0947 & 18.74 & 7.81 & 7.86 &  43.86  &  Yes & this work \\ 
J155259.94$+210246.8$ & 238.24977 & $21.04638$ & 0.1717 & 19.12 & 14.39 & 8.44 &  59.24  &  Yes & this work, LN23b \\ 
J160808.21$+161336.5$ & 242.03421 & $16.22688$ & 0.1100 & 18.96 & 8.77 & 17.64 &  121.41  &  \nodata & \nodata \\ 
J161219.56$+462942.5$ & 243.08154 & $46.49517$ & 0.1254 & 18.53 & 8.30 & 2.24 &  26.05  &  Yes & this work, LN23b \\ 
J163639.57$+194201.7$ & 249.16492 & $19.70048$ & 0.1496 & 18.99 & 5.68 & 0.69 &  12.71  &  No & this work \\ 
J164210.34$+445733.1$ & 250.54312 & $44.95922$ & 0.2224 & 20.11 & 5.35 & 0.02 &  6.20  &  \nodata& \nodata \\
J215055.73$-010654.1$ & 327.73221 & $-1.11502$ & 0.0879 & 19.93 & 8.64 & 2.92 &  27.35  &  Yes & LN22  \\ 
J222559.66$+201944.7$ & 336.49863 & $20.32912$ & 0.1554 & 18.85 & 12.57 & 4.77 &  46.08  &  Yes & this work \\ 
J232257.70$+152757.6$ & 350.74045 & $15.46599$ & 0.0433 & 17.84 & 5.52 & 8.94 &  36.41  &  \nodata & \nodata \\  \hline
\end{tabular}
\end{center}
\end{table*}

\begin{table*}[htp]
\caption{Turn-off CL~AGN candidates}\label{tab:turn-off-candidates}
\begin{center}
\begin{tabular}{ c c c c c c c c c c c }
\hline \hline
   SDSS identifier & RA & DEC & $z$ & $g_{\rm MED}$ & $\overline{g_{\rm err}}$ & $\sigma_{\rm QSO}$ & $\sigma_{\rm NotQSO}$ & $\sigma_{\rm var}$   & Confirmed? & Reference \\ 
                               &  degree & degree  &    &  mag &  mag &   &   &   &  & \\ \hline 
J023254.67$-045050.9$ & 38.22777 & $-4.84746$ & 0.2060 & 19.81 & 0.12 & 0.63 & 4.01 & 8.85 &  \nodata & \nodata \\
J023259.52$-051145.7$ & 38.24802 & $-5.19602$ & 0.1420 & 19.46 & 0.09 & 0.89 & 1.23 & 4.71 &  \nodata & \nodata \\ 
J082615.60$+405806.2$ & 126.56498 & $40.96840$ & 0.3219 & 19.92 & 0.13 & 0.76 & 0.19 & 1.88 &  \nodata  & \nodata \\ 
J085846.56$+274534.5$ & 134.69398 & $27.75958$ & 0.2890 & 19.80 & 0.10 & 0.93 & 2.64 & 6.98 &  \nodata & \nodata \\
J101034.28$+372514.9$ & 152.64285 & $37.42080$ & 0.2824 & 19.11 & 0.07 & 0.49 & 0.00 & 0.05 &  \nodata  & \nodata \\ 
J102152.36$+464515.7$ & 155.46815 & $46.75437$ & 0.2040 & 19.50 & 0.10 & 0.85 & 1.21 & 4.12 &  Yes & M19 \\ 
J112129.07$+475509.9$ & 170.37114 & $47.91943$ & 0.3160 & 19.96 & 0.15 & 0.45 & 2.92 & 6.56 &  \nodata & \nodata \\ 
J123020.67$+462745.8$ & 187.58614 & $46.46273$ & 0.1803 & 19.08 & 0.08 & 0.82 & 1.16 & 3.74 &  Yes &  this work \\ 
J131211.54$+123707.0$ & 198.04807 & $12.61860$ & 0.3063 & 19.23 & 0.07 & 0.58 & 0.42 & 1.68 &  \nodata  & \nodata \\ 
J133241.18$+300106.7$ & 203.17157 & $30.01854$ & 0.2204 & 19.82 & 0.11 & 0.48 & 1.72 & 4.50 &  Yes & this work \\ 
J140007.94$+275133.3$ & 210.03307 & $27.85926$ & 0.3338 & 19.02 & 0.07 & 0.96 & 0.04 & 0.60 &  No  & this work \\ 
J143455.32$+572345.3$ & 218.73048 & $57.39591$ & 0.1752 & 19.34 & 0.09 & 0.78 & 0.00 & 0.00 &  Yes & M19 \\  
\hline
\end{tabular}
\end{center}
\end{table*}

\section{Spectral decomposition} \label{sec:decomposition}

The decomposition was done in the rest-frame after correcting Galactic extinction using the extinction curve of \citet{Cardelli89} and dust map of \citet{Schlegel98}.  It adopt a Principle Component Analysis (PCA) to decompose the host galaxy and quasar component  using the eigenspectra from \citet{Yip04a, Yip04b}.  After subtracting the host component, the residual spectra were fitted with a global pseudo-continuum model using several emission-line free windows \citep{Shen19} within [3800\AA, 7500\AA].  The pseudo-continuum model consists of a power-law component and a \FeII\ template  \citep{Boroson92} with three parameters to scale, shift and broaden it.   After subtracting the best-fit continuum model,  we fit \hbeta\ and \halpha\ line complex separately. In \hbeta\ complex, we include narrow and broad  \hbeta\, central and wing \OIIIa\ and \OIIIb\,  narrow and broad \HeIIopt. in \halpha\ complex, we include narrow and broad \halpha, narrow \NIIab, and narrow \SIIab. For each narrow component, we use one single Gaussian. For broad component, we allow maximum three Gaussians to better describe the asymmetry if any.

We chose to fit bright and dim state separately rather than taking the host model from dim state as a prior for bright state, because the latter approach sometimes cannot fit the bright state well \citep{Green22}, which is possibly due to different seeing/weather conditions among different epochs.  Figure \ref{fig:decomposition} displays an example of our fit to both faint and bright state of an individual object.  We should note that, our candidates are all Type 1.9/2 AGNs in their dim state, so their AGN continuum at faint state are not well constrained. However, we only use the line luminosity (flux) in the following analysis, so the detailed host and continuum subtraction will not affect our result.

We measured the line luminosity and the FWHM for \hbeta\ and \halpha\ using the sum of the broad Gaussians. The uncertainties of these quantities are measured using Monte-Carlo approach where we generate 50 mock spectra by perturb the original spectra using its error spectra. The final measurement and its uncertainty are taken as the average and 16-to-84 percentile of the posterior distribution.

\setcounter{figure}{0}   

\begin{figure*}[htbp]
    \centering
    \includegraphics[width=0.98\textwidth]{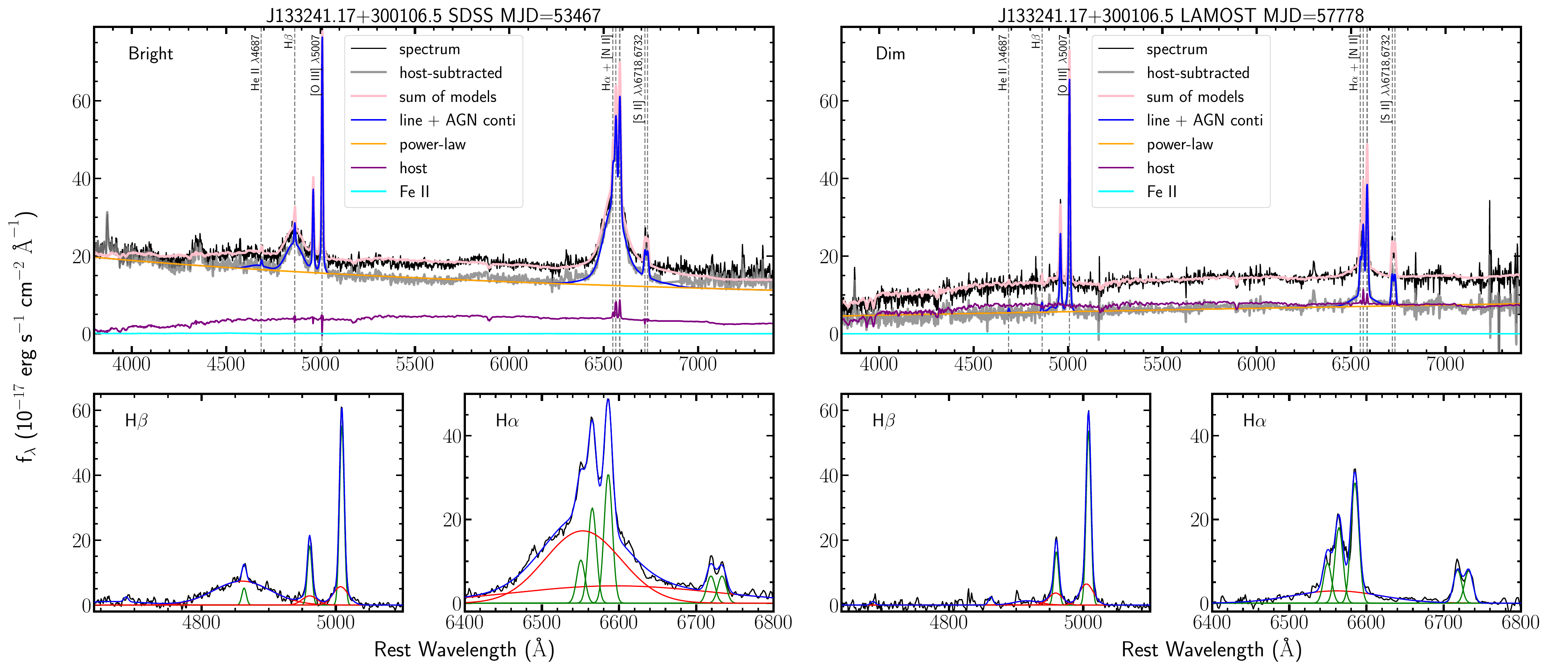}
 
    \caption{Example of spectral decomposition for bright (left) and dim (right) state of J133241.18$+$300106.7 using {\tt PyQSOFit}.  The upper panels exhibit the full modeling results, where the original spectra, host galaxy component, host subtracted spectrum, AGN power-law continuum and \FeII\ component are shown in black, purple, grey, orange and cyan respectively. The blue solid lines represent the sum of emission-line models. The lower panels zoom in to the  \hbeta\ (left) and \halpha\ (right) region to display the line fit result. The green and red lines represent individual narrow and broad Gaussians, respectively.   }
    \label{fig:decomposition}
\end{figure*}

\vspace{5mm}

\bibliography{ref.bib}\label{sec:ref}
\bibliographystyle{aasjournal}

\end{document}